\documentclass[useAMS,usenatbib,fleqn,a4paper]{mn2e}
\usepackage{times}
\usepackage{amsmath}
\usepackage{amssymb}
\usepackage{amsbsy}
\usepackage{bmpsize}
\usepackage{graphicx}
\usepackage{subfigure}
\usepackage{paralist}
\usepackage{mathrsfs}
\usepackage{booktabs}
\usepackage{tabularx}

\def\kpc{\,{\rm kpc}}

\def\pc{\,{\rm pc}}

\def\kms{\,{\rm km\,s^{-1}}}
\def\rad{\,\rm rad}
\def\Gyr{\,{\rm Gyr}}
\def\Myr{\,{\rm Myr}}

\def\percent{\text{ per cent}}

\newcommand{\bs}[1]{\bmath{#1}}

\def\percent{\text{ per cent}}

\usepackage{color}
\definecolor{darkred}{rgb}{0.55, 0.0, 0.0}
\definecolor{darkblue}{rgb}{0.0, 0.0, 0.55}
 \definecolor{darkgreen}{rgb}{0.0, 0.2, 0.13}
\usepackage[colorlinks=true,linkcolor=darkred,citecolor=darkblue,urlcolor=darkgreen]{hyperref}

\title{Dynamics of stream-subhalo interactions}
\author[J. L. Sanders, J. Bovy and D. Erkal]{Jason L. Sanders$^1$\thanks{E-mail: jls@ast.cam.ac.uk}, Jo Bovy$^2$ and Denis Erkal$^1$\\
$^1$Institute of Astronomy, Madingley Road, Cambridge, CB3 0HA\\
$^2$Department of Astronomy and Astrophysics, University of Toronto, 50 St. George Street, Toronto, ON M5S 3H4}
\pagerange{\pageref{firstpage}--\pageref{lastpage}} \pubyear{2015}

\begin{document}
\maketitle
\label{firstpage}
\begin{abstract}
We develop a formalism for modelling the impact of dark matter subhaloes on cold thin streams. Our formalism models the formation of a gap in a stream in angle-frequency space and is able to handle general stream and impact geometry. We analyse an $N$-body simulation of a cold stream formed from a progenitor on an eccentric orbit in an axisymmetric potential, which is perturbed by a direct impact from a $10^8 M_\odot$ subhalo, and produce a complete generative model of the perturbed stream that matches the simulation well at a range of times. We show how the results in angle-frequency space can be related to physical properties of the gaps and that previous results for more constrained simulations are recovered. We demonstrate how our results are dependent upon the mass of the subhalo and the location of the impact along the stream. We find that gaps formed far downstream grow more rapidly than those closer to the progenitor due to the more ordered nature of the stream members far from the progenitor. Additionally, we show that the minimum gap density plateaus in time at a value that decreases with increasing subhalo mass.
\end{abstract}


\section{Introduction}

One of the key predictions of the currently favoured $\Lambda$CDM cosmology is hierarchical structure formation. Early in the Universe dark matter overdensities formed, began accumulating baryons, and conglomerated into ever larger dark matter haloes. The result of this process is that at the current time there are many smaller dark matter haloes that are orbiting within a large dark matter halo such as that of the Milky Way. Current dark-matter-only simulations \citep[e.g.][]{Diemand2008,Springel2008} make predictions for the slope of the mass spectrum of these subhaloes, although the exact structure of the subhalo mass spectrum in $\Lambda$CDM is still an ongoing area of research \citep[e.g.][]{Giocoli2010,Gao2011}. Warm dark matter models predict quite different subhalo mass spectra \citep{Lovell2014} such that the presence and number of these low mass subhaloes is a crucial test for the $\Lambda$CDM cosmology.

From an observational point-of-view the nature of dark matter has remained aloof as dark matter has not yet been directly detected although the direct detection experiments are ruling out possible dark matter candidates \citep{Bertone2005,Feng2010}. The other crucial line of attack is probing the gravitational effects of dark matter on both visible matter and photons on astrophysical scales. The large-scale smooth dark matter halo is being mapped out with dynamical models of both the assumed phase-mixed components of the Galaxy e.g. the disc \citep[e.g.][]{BovyRix2013, Piffl2014}, and those components of the Galaxy that are yet to fully phase mix, e.g. tidal streams \citep[e.g.][]{Koposov2010,Gibbons2014,Kuepper2015,Bowden2015}. Tidal streams are filaments of material stripped from satellites of a host galaxy and are naturally a result of the hierarchical structure formation picture. Some streams (e.g. GD-1, Palomar 5) are observed to be very kinematically cold such that, in addition to being probes of the large-scale dark matter structure, they are very sensitive probes of interactions with the dark matter subhaloes. Interactions between streams and dark matter haloes produce gaps in the streams that grow over time. A conclusive detection of one such gap induced by a low-mass subhalo would be very powerful confirmation of $\Lambda$CDM cosmology, whilst the ultimate goal might be to measure the dark-matter halo mass spectrum from a whole series of observed gaps.

To achieve this ambitious goal we need a concerted modelling effort to understand the structure of a gap in a stream. The goal is doubly ambitious due to the difficulties that modelling unperturbed streams has presented. The modelling of streams has received considerable attention in the literature and there have been many proposed methodologies for measuring the Galactic potential using streams. However, there have been relatively few successful applications to the data \citep[e.g.][]{Koposov2010,LawMajewski2010,Gibbons2014,Kuepper2015,Bowden2015} and we have not yet reached a point where modelling a stream is simple. One very powerful method is to construct models of streams in angle-frequency space as proposed by \cite{Bovy2014} and \cite{Sanders2014}. This space simplifies the stream dynamics \citep{HelmiWhite1999,Tremaine1999} and allows rapid generation of general stream models.

In recent years there has been much progress made in understanding the structure of stream gaps through the analysis of numerical simulations \citep{Johnston2002,Ibata2002,SiegalGaskins2008,Carlberg2009,YoonJohnstonHogg,Carlberg2012} as well as through analytic approaches \citep{Carlberg2013,ErkalBelokurov2015,ErkalBelokurov2015b}. \cite{ErkalBelokurov2015} extended the treatment introduced in \cite{Carlberg2013} and developed a simple picture of how gaps form in streams on circular orbits due to changes in the orbital frequency induced by a subhalo's passage. The analytic results of that work were then used in \citet{ErkalBelokurov2015b} to demonstrate that subhalo properties can be reliably inferred from observations of streams with realistic observational errors. While this simple picture of gap formation has provided useful insight, it relies on the ease with which nearly circular orbits can be handled and cannot be easily extended to realistic streams which are on eccentric orbits.

Motivated by these analytic results, we will build on the angle-frequency stream formalism and argue that it provides a clear basis on which to incorporate perturbations due to subhalo fly-bys. We show that many of the insights from circular orbits in \cite{ErkalBelokurov2015} can be extended to general, eccentric orbits. We work from the point-of-view that we know the underlying unperturbed stream distribution and the Galactic potential well and we are trying to characterise substructure in the stream.

We will show how one can compute the kicks in velocity, angles and frequencies for general subhaloes and general impact geometries. With the aid of an $N$-body simulation of a cold stream on an eccentric orbit in a flattened axisymmetric logarithmic potential we demonstrate how a perturbed stream model can be generated and we inspect the resultant distributions in action, angle and frequency space. We show that a gap forms in angle, frequency and action space and that the gap size in the parallel angle space (the angle along the stream) grows like the spatial gap size. We simulate a series of halo fly-bys of differing masses and geometry. We find in all cases the minimum density in the gap plateaus in time and that the gap size grows fastest for fly-bys far from the stream progenitor where the particles are well ordered by energy. Finally, we show that we are able to produce a fully generative model of a perturbed stream that matches the $N$-body simulation well in Galactocentric coordinates.

The paper is arranged as followed. We begin in Section~\ref{Sect::Formalism} with the framework for computing the velocity perturbations of a stream due to a general subhalo fly-by. We compute the velocity perturbations using methods of increasing complexity, and show the difference between velocity kicks computed for a range of subhalo profiles. In Section~\ref{Sec::Formalism_angfreq} we translate these velocity perturbations into angle and frequency perturbations and present the formalism for perturbing an angle-frequency stream model. In Section~\ref{Sect::Simulation} we detail an $N$-body simulation of a $10^8M_\odot$ dark matter impact on a cold stream formed from a progenitor on an eccentric orbit. We project the simulation into action, angle and frequency coordinates and by applying the framework of Section~\ref{Sect::Formalism} we perturb an unperturbed simulation snapshot and compare to the perturbed simulation snapshot. We show the angle and frequency kicks that result from a subhalo fly-by and develop an analytic approximation that well reproduces the frequency kicks. To close the section we compute the gap size in angle space as a function of time and by projecting the models back into real space show how the gap size in angle space correlates with the spatial gap in the stream.  In Section~\ref{Sect::Variation} we consider fly-bys of varying subhalo masses and varying impact geometry. In Section~\ref{Sect::FULL} we demonstrate how a fully generative perturbed stream model can be created and show the model density as well as the configuration space distributions well match the simulation. We present improvements to the unperturbed stream model that are necessary to reproduce the underlying stream density. We discuss the applicability of our formalism and present our conclusions in Section~\ref{Sect::Conclusions}.

\section{Velocity Perturbations}\label{Sect::Formalism}
In this section we give expressions for the velocity changes of stream particles under the influence of a subhalo fly-by. Following \cite{YoonJohnstonHogg}, \cite{Carlberg2013} and \cite{ErkalBelokurov2015} we work under the impulse approximation i.e. the integrated acceleration is assumed to act instantaneously at some impact time such that the velocity of the stream particles changes instantly. The change in velocity is given by
\begin{equation}
\delta\bs{v}^g = \int_{-\infty}^{\infty} \mathrm{d} t\,\bs{a}(\bs{x}(t)),
\end{equation}
where $\bs{a}$ is the acceleration resulting from the force exerted by the subhalo which depends on the location $\bs{x}$ of the particle at time $t$\footnote{Throughout this paper we use $\delta$ to denote changes in quantities for individual particles in time e.g. from the subhalo fly-bys, and $\Delta$ for differences \emph{between} particles.}. The fly-by is described by the impact parameter $b$, the velocity of the subhalo $\bs{w}$ and the time of the impact $t=-t_g$. To simplify the expressions in this section, we set $t_g=0$.

\cite{ErkalBelokurov2015} gave expressions for the velocity kicks due to a Plummer subhalo on a straight stream segment moving at a fixed relative velocity to the subhalo. If we consider the example of GD-1 and take the model of \cite{Koposov2010} we find that the observed segment of the stream has a radius of curvature of $\sim24\kpc$. \cite{ErkalBelokurov2015} show that for a $10^8M_\odot$ Plummer subhalo with scale radius $r_s=625\pc$ the kicks are important on scales of $\sim10r_s\approx6\kpc$ such that the radius of curvature is comparable to the region over which the kick is important. Similarly, from \cite{ErkalBelokurov2015} we know that the typical distance a stream particle moves during the interaction is $\sim10|\bs{v}|r_s/|\bs{w}-\bs{v}|\approx10r_s/\sqrt{2}\approx4\kpc$ (where $\bs{w}$ is the velocity of the subhalo) such that the curvature of the orbit of an individual particle is also important over the interaction. Here we will give more general formulae for the velocity kicks that account for the curved extent of the stream.

\subsection{Plummer subhalo kicks on curved stream}

\begin{figure}
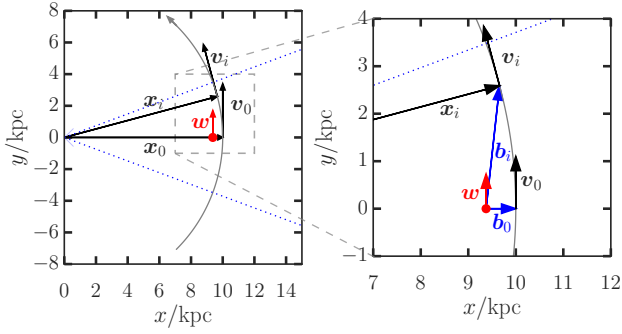

$$\includegraphics[width=\columnwidth]{{{plots/fig1_geometry}}}$$
\caption{Stream-subhalo interaction geometry at the point of closest approach: the right panel is a zoom-in of the region inside the grey dashed square in the left panel. The stream track is a circular orbit of radius $10\kpc$ and circular velocity $220\kms$ shown by the grey line. The red vector $\bs{w}$ shows the projection of the sub-halo velocity. The black position vectors $\bs{x}_0$ and $\bs{x}_i$ show the positions of the stream closest to the subhalo and an arbitrary stream point respectively. The black velocity vectors $\bs{v}_0$ and $\bs{v}_i$ give the velocities of these two stream points. The blue vectors $\bs{b}_0$ and $\bs{b}_i$ are the closest-approach displacement vector and the displacement vector between the subhalo and the arbitrary stream point respectively. The dotted blue lines show the angles at which the projection of the velocity vector of the subhalo intersects the stream.
}
\label{geometry_lb}
\end{figure}

We expand on the formalism of \cite{ErkalBelokurov2015} by explicitly considering the spatial distribution of the stream. We retain the simplification that each particle moves at a fixed relative velocity to the subhalo during the fly-by. We define the phase-space coordinates of the stream particles as $(\bs{x}_i,\bs{v}_i)$ and the stream point of closest approach as $(\bs{x}_0,\bs{v}_0)$. The geometry of the interaction is shown in Fig.~\ref{geometry_lb}. The vector of closest approach is given by
\begin{equation}
\bs{b}_0=b\frac{\bs{w}\times\bs{v}_0}{|\bs{w}\times\bs{v}_0|}.
\end{equation}
Note the sign of $b$ is important in defining the curvature of the stream relative to the subhalo. We define
\begin{equation}
\bs{b}_i = \bs{b}_0+\bs{x}_i-\bs{x}_0
\end{equation}
as the displacement vector between stream particle $i$ and the subhalo at the impact time, and
\begin{equation}
\bs{w}_i=\bs{w}-\bs{v}_i
\end{equation}
as the corresponding relative velocity. The expression for the velocity kicks for the $i$th stream particle is given by

\begin{equation}
\begin{split}
\delta\bs{v}^g_i &= \int_{-\infty}^{\infty} \mathrm{d} t\,\bs{a}(\bs{x}_i(t)),\\
\bs{x}_i(t)&=\bs{b}_i+\bs{w}_i\,t.
\end{split}
\end{equation}

For general subhalo acceleration fields this integral must be calculated numerically via a coordinate transformation to make the limits finite. However, for a Plummer subhalo the velocity kicks may be computed analytically. The potential of a Plummer sphere is given by
\begin{equation}
\Phi_{\rm P}(r) = -\frac{GM}{\sqrt{r^2+r_s^2}},
\end{equation}
where $M$ is the mass and $r_s$ the scale radius. The velocity kicks are given by
\begin{equation}
\delta\bs{v}^{g}_{i,\mathrm{P}} = -\frac{2GM}{|\bs{w}_i|}
\frac{\bs{b}_i-\hat{\bs{w}}_i(\bs{b}_i\cdot\hat{\bs{w}}_i)}
{(B^2+r_s^2)},
\label{curved}
\end{equation}
where
\begin{equation}
B^2=|\bs{b}_i|^2
-|\bs{b}_i\cdot\hat{\bs{w}}_i|^2.
\end{equation}

\subsection{Comparison of methods for computing the velocity kicks}\label{diff_methods}
\begin{figure*}
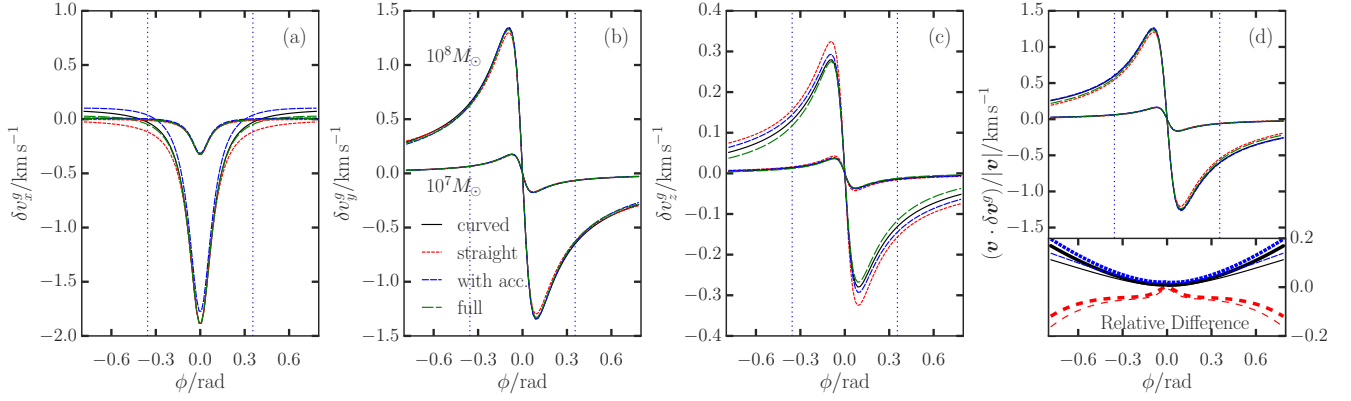

$$\includegraphics[width=\textwidth]{{{plots/fig2_curved_straight_comparison}}}$$
\caption{
A comparison of the velocity kicks calculated with varying levels of complexity for the interaction between a Plummer sphere and a circular stream track as described in Section~\ref{diff_methods}. The \textbf{black lines} show the kicks calculated using equation~\protect\eqref{curved} (i.e. fixed relative velocity during fly-by), whilst the \textbf{red short-dashed lines} show the kicks computed assuming the stream is a straight line segment along the $y$-axis, the \textbf{blue medium-dashed lines} shows the kicks computed using the full path of each particle during the fly-by and the \textbf{green long-dashed lines} show the kicks computed by integrating each particle in the combined galactic and orbiting Plummer potential. The dotted blue lines show the angles at which the projection of the velocity vector of the subhalo intersects the stream. Each panel shows two lines corresponding to two different subhalo mass impacts of $10^8M_\odot$ and $10^7M_\odot$. \textbf{Panels (a), (b) \& (c)}: The first, second and third plots show the Cartesian velocity kicks as a function of the azimuthal angle from the centre of the galaxy. \textbf{Panel (d)}: The rightmost panel shows $\bs{v}\cdot\delta\bs{v}^g$ with the small inset showing the relative difference between the lines. The black lines show the difference between the full computation and the kicks computed with equation~\protect\eqref{curved}, the red short-dashed lines show the difference between the full computation and the kicks computed assuming the stream is a straight line, and the blue long-dashed show the difference between the full computation and that including the acceleration of each particle during the fly-by. The thicker lines correspond to the $10^8M_\odot$ subhalo.
}
\label{curved_straight}
\end{figure*}

In Fig.~\ref{curved_straight} we show a comparison between the kicks calculated using approaches of increasing complexity. We compute the kicks calculated assuming the stream is a straight-line segment as in \cite{ErkalBelokurov2015}, the kicks calculated using the curvature of the stream track as in equation~\eqref{curved}, the kicks calculated using the full orbital path of each stream particle but assuming the subhalo moves in a straight line, and finally the kicks calculated using the full orbit of the stream particles \emph{and} the subhalo. The last of these is computed by first integrating each particle and the subhalo backwards in time for some time $T$ in the galactic potential, integrating the particles forward in the combined galactic and subhalo potentials for a time $2T$ and finally integrating the particles backwards for a time $T$ and computing the difference between the initial and final velocities.

The particles are evolved in a flattened logarithmic potential of the form
\begin{equation}
\Phi_\mathrm{L}(R,z) = \frac{V_c^2}{2}\log\Big(R^2+\frac{z^2}{q^2}\Big),
\label{potential}
\end{equation}
with $V_c = 220\kms$ and $q = 0.9$ \citep{Koposov2010}, and the stream here is modelled simply as a circular orbit at $10\kpc$ with a circular velocity of $220\kms$ impacted by a Plummer sphere with velocity $\bs{w}=(0,132,352)\kms$. The choice of velocity is motivated by the work of \cite{PifflBinney} who constrained the smooth dark matter distribution to have a velocity dispersion at the Sun of $\sigma_i\approx150\kms$ and we assume that the dark matter subhalo distribution function is similar to the smooth dark matter distribution. Additionally, the velocity was chosen such that the subhalo only interacts with the stream once over the progenitor's orbital period. The impact parameter is $b=-625\pc$ where the minus sign denotes that the subhalo passes inside the circular orbit of the progenitor. The distance along the stream in the straight-line approximation is calculated as $\Delta\phi\times10\kpc$.

We show two kicks: one due to a subhalo of mass $10^8M_\odot$ and scale radius $r_s=625\pc$ and one due to a subhalo of mass $10^7M_\odot$ and scale radius $r_s=250\pc$. The parameters were chosen such that the subhaloes lie within the maximum circular velocity against tidal mass (approximating the tidal mass as the total mass) relation from the Via Lactea II catalogue of \cite{Diemand2008}. The maximum velocity kick from a direct Plummer subhalo impact is proportional to the maximum circular velocity \citep{ErkalBelokurov2015} so matching the maximum circular velocity gives realistic velocity kicks. Our chosen subhalo parameters lie above the average relation from \cite{Diemand2008} as they are more concentrated. However, the considered impacts are physically plausible as the subhaloes still lie within the scatter of the Via Lactea II distributions.

The differences between the different methods are small. One very noticeable difference is the amplitude of the $\delta v^g_x$ kicks computed using the orbital path of the stream particles, but with the subhalo moving in a straight line, is significantly smaller than for the other three cases. This is because the stream particles curve towards the subhalo track and so experience an increased perturbation. This picture is confirmed by moving the subhalo fly-by to outside the stream track which produces an under-estimate for $|\delta v^g_x|$. Additionally, the $\delta v^g_z$ kicks for the full stream and subhalo integration are smaller at large distances from the closest approach. We also show $\bs{v}\cdot\delta\bs{v}^g$ which is very similar for the four cases. This quantity is the change in energy $\delta H$ and is related by a constant scaling to $\Delta E/\Delta E_\mathrm{char}$ plotted in Figure 4 of \cite{YoonJohnstonHogg}. For many potentials the orbital frequencies are strongly dependent on the energy and only weakly dependent on the other integrals of motion \citep{BinneyTremaine}. Therefore it is this quantity which has the largest effect on the future structure of the stream.  Additionally, \cite{ErkalBelokurov2015} argued that for the near-circular orbit case the azimuthal velocity kicks and hence $\bs{v}\cdot\delta\bs{v}^g\approx v_\phi\delta v_\phi^g$ produced the most significant effects on the gap evolution. In the bottom inset panel we show the relative difference in $\bs{v}\cdot\delta\bs{v}^g$ between the three approximate methods and the full orbit integration method. For both subhaloes the straight line approximation produces a similar relative difference with the $10^8M_\odot$ case being slightly more accurate. For the other two cases (`curved' and `with acc.') the $10^7M_\odot$ case produces a smaller relative difference (approximately $5\percent$ smaller for $|\phi|\gtrsim0.6\rad$ ). It seems that if we account for the curvature of the stream track the $10^7M_\odot$ interaction is better approximated as impulsive.

In conclusion we find that all the absolute differences between the velocity kicks using different methods are very small, particularly for the $10^7M_\odot$ impact, and probably much smaller than the precision with which line-of-sight velocities will be measured for a number of years. Therefore, we conclude that the straight-line approximation is sufficient. However, the curved approximation has the same computational cost, whilst capturing an additional physical effect, so for our calculations in Section~\ref{Sect::Simulation} we will use equation~\eqref{curved} to calculate all kicks. We note however that we have not explored different impact and stream geometries but rather inspected what we see as a representative case so there may be special cases where the more computationally expensive methods are required.

\subsection{Different subhalo profiles}

\begin{figure*}
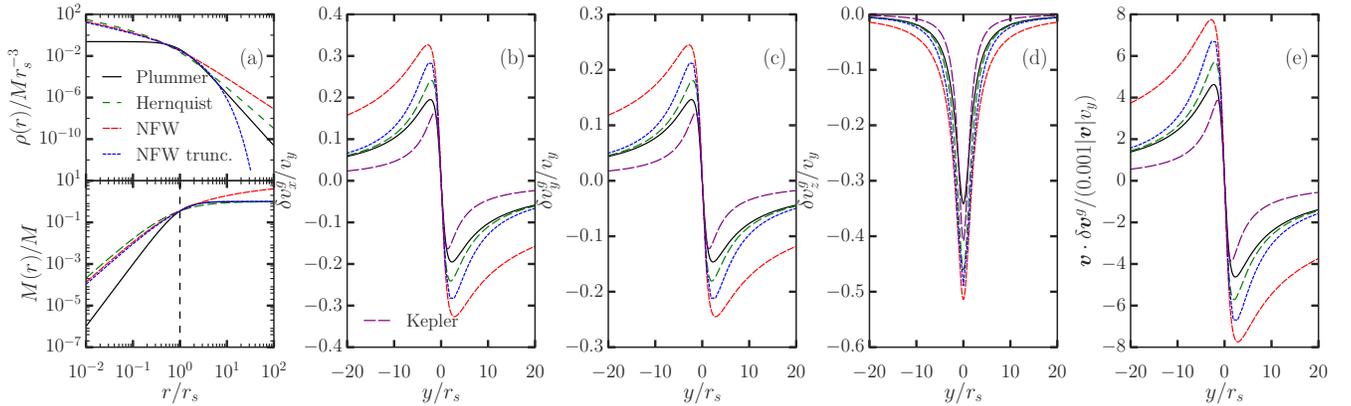

$$\includegraphics[width=\textwidth]{{{plots/fig3_nfw_plummer_comparison}}}$$
\caption{
A comparison of the velocity kicks from a Plummer (black solid), a Hernquist (green long-dashed), a NFW (red medium-dashed) and a truncated NFW subhalo (blue short-dashed). \textbf{Panel (a)}: The left panel shows the density (top) and mass (bottom) profiles of the four haloes. The three finite mass haloes have the same total mass and the NFW profile has the same mass parameter. The Hernquist, NFW and truncated NFW scale radii are chosen such that at the Plummer scale radius ($r_s$) the same mass is enclosed by the profiles. The truncated NFW halo has a truncation radius equal to twice its scale radius. \textbf{Panels (b), (c) \& (d)}: The second, third and fourth panels show the velocity kicks as a function of the distance along the stream.  \textbf{Panel (e)}: The rightmost panel shows $\bs{v}\cdot\delta\bs{v}^g$. The stream is a particle train along the $y$-axis with velocity $v_y$. The subhaloes have $4GM=r_s v_y^2$, $r_{s,\mathrm{Hernquist}}=1.36r_s$, $r_{s,\mathrm{NFW}}=1.21r_s$, $r_{s,\mathrm{NFW trunc}}=1.76r_s$, impact parameter $b=\sqrt{3}/2r_s$ and velocity $\bs{w}=(0.3,0.6,0)v_y$. In the second through fifth panel we show in long-dashed purple the kicks from a point-mass (or Kepler potential) with mass equal to the enclosed mass of the Plummer halo at the scale radius.
}
\label{plummer_nfw}
\end{figure*}

We now compare the kicks produced by a Plummer profile with those due to other astrophysically interesting profiles for which we compute the kicks numerically. We consider kicks due to Hernquist profiles \citep{Hernquist} and truncated Navarro-Frenk-White \citep[NFW,][]{NFW1996} profiles. The density profile for the truncated NFW halo is given by
\begin{equation}
\rho_{\rm NFW}(r) = \frac{M}{4\pi r_s^3}\Big(\frac{r}{r_s}\Big)^{-1}\Big(1+\frac{r}{r_s}\Big)^{-2}\mathrm{sech}\Big(\frac{r}{r_t}\Big).
\end{equation}
$r_t$ is the truncation radius. For non-zero $r_t$ the forces and potential must be computed numerically and clearly for $r_t\rightarrow\infty$ the form reduces to the well-known NFW profile. \cite{Hayashi2003} find that tidally-stripped dark-matter haloes have a polynomial truncation of $\rho\propto r^{-6}$ at large radii so the exponential truncation should be treated as the most extreme truncation and more realistic NFW halos will lie somewhere between our truncated and non-truncated lines.

In Fig.~\ref{plummer_nfw} we show the kicks computed for a Plummer subhalo, a Hernquist subhalo, an NFW subhalo and truncated NFW subhalo. We also show $\bs{v}\cdot\delta\bs{v}^g$ which is the quantity that controls the future structure of the stream. We have modelled the stream as a straight-line segment moving along the $y$ axis at velocity $v_y$. The Plummer subhalo has a scale-radius $r_s$ and satisfies $4GM=r_sv_y^2$. The other two finite-mass haloes have the same total mass as the Plummer sphere, with the NFW halo mass parameter chosen to be equal to the total mass of the Plummer sphere. All haloes have scale radii chosen such that the mass contained within $r_s$ is identical to the enclosed mass for the Plummer subhalo. This results in scale radii given by $r_{s,\mathrm{Hernquist}}=1.36r_s$, $r_{s,\mathrm{NFW}}=1.21r_s$, $r_{s,\mathrm{NFW trunc}}=1.76r_s$. The truncated NFW profile has a truncation radius of $r_t=2r_{s,\mathrm{NFW trunc}}$. We also show, for reference, the kicks from a point-mass, or Kepler potential, with mass equal to the mass enclosed by the Plummer sphere at $r_s$. The subhaloes have velocity $\bs{w}=(0.3,0.6,0)v_y$ and impact parameter $b=\sqrt{3}/2r_s$.

Of the spatially-extended haloes, the Plummer sphere produces the smallest kicks at all radii as it has the least mass at small radii. The Hernquist sphere produces the second smallest kicks as the mass enclosed for $r_s<r<10r_s$ is smallest for the Hernquist sphere. The NFW and truncated NFW profiles produce similar amplitude kicks at small distances but the truncated NFW kicks fall off faster at larger distances. The NFW profile is the most spatially extended so it produces the largest kicks at large distances. However, such a dark matter subhalo is unphysical so we only include it here for completeness. The three finite-mass spatially-extended subhaloes produce kicks that tend towards each other at large distances. The Kepler potential produces smaller kicks than the spatially-extended haloes due to its lower total mass.

\section{Angle-frequency perturbations}\label{Sec::Formalism_angfreq}
\cite{ErkalBelokurov2015} developed a simple model for gap formation and gave analytic results for the structure of a gap as a function of time for streams on circular orbits. Such an approach is fruitful for developing an understanding of the gap formation but when modelling realistic streams we require a formalism that is appropriate for eccentric orbits.

Dynamical systems are often simplified through the use of angle-actions coordinates $(\bs{J},\btheta)$ (see \citealt{BinneyTremaine}). These canonical coordinates possess the properties that the actions are integrals of motion whilst the angles increase linearly with time at a constant rate $\boldsymbol{\Omega}$, which are the frequencies. The equations of motion for the angle-action coordinates are
\begin{equation}
\begin{split}
\bs{J}&=\mathrm{constant},\\
\btheta&=\frac{\partial H}{\partial\boldsymbol{J}}t+\btheta(0)=\boldsymbol{\Omega}t+\btheta(0),
\end{split}
\label{angleactdef}
\end{equation}
where $t$ is the time and $\btheta(0)$ is the angle at $t=0$. In axisymmetric potentials\footnote{In this paper we will only work with axisymmetric potentials but the formalism is simply extended to more general potentials.} the three actions are given by $\bs{J}=(J_R,J_\phi,J_z)$. $J_R$ is the radial action describing the extent of the radial oscillations, $J_z$ is the vertical action describing the extent of the vertical oscillations and $J_\phi$ is the $z$-component of the angular momentum. The frequencies $\boldsymbol{\Omega}=(\Omega_R,\Omega_\phi,\Omega_z)$  are the corresponding rates of the oscillations with the angles $\boldsymbol{\theta}=(\theta_R,\theta_\phi,\theta_z)$ describing the phase of the oscillation.

The meaning of the angle-action coordinates can be demonstrated by considering circular orbits. In this case the angle $\theta_\phi$ is the azimuthal angle $\phi$ around the orbit and the frequency $\Omega_\phi$ is the circular frequency. For near-circular orbits, the epicyclic oscillations about the circular orbit are described by the two frequencies $\Omega_R$ and $\Omega_z$. The angle $\theta_z$ describes the oscillation phase perpendicular to the orbital plane, whilst $\theta_R$ and $\theta_\phi-\phi$ are dependent and describe the circulation about the guiding centre \citep[][equation 3.265]{BinneyTremaine}
. \cite{ErkalBelokurov2015} considered the case of near-circular orbits when analysing the structure of a gap and we will see that many of the results for more general orbits can be related back to the near-circular orbit case, so it is useful to keep in mind the meaning of the angle and frequency variables for the near-circular case.

To compute the actions, angles and frequencies in this paper we use the method detailed by both \cite{SandersBinney2014} and \cite{Bovy2014}\footnote{The two cited methods differ in the action computation as \cite{Bovy2014} averages the toy actions over the toy angles whilst \cite{SandersBinney2014} solve for the Fourier coefficients.}. This method uses a generating function to transform from a set of toy angle-action variables (computed in an isochrone potential) to the target set. The coefficients of the generating function (Fourier components $S_{\bs{n}}$ and their derivatives) are found from a least-squares fit to a series of phase-space samples from an orbit integration. Note that this method is not limited to axisymmetric potentials and can be used for more general static potentials e.g. triaxial potentials \citep[see][]{SandersBinney2014}. For the orbit of the stream progenitor considered in this paper the actions are computed with a relative accuracy of $3\times10^{-6}$, the frequencies to a relative accuracy of $3\times10^{-5}$ and the $\theta_i/\pi$ are computed to an absolute accuracy of $1\times10^{-4}$ (these quantities are computed by measuring the standard deviation of the action and frequency estimates, and the standard deviation of the angle estimates about a straight line, for a series of samples from a $300\Gyr$ orbital segment).

The angle-action coordinates simplify the description of the evolution of a tidal stream, as once stripped the stream members move essentially as free particles in the galactic potential. \cite{HelmiWhite1999} and \cite{Tremaine1999} both discussed the evolution of a stream as a small action clump in angle-actions, which was further built on by the work of \cite{EyreBinney2011}. \cite{SandersBinney2013b} presented the idea that the angle-frequency distribution could be used as a probe of the galactic potential. These ideas were extended by both \cite{Bovy2014} and \cite{Sanders2014} who developed generative models for streams in angle-frequency space. This framework is ideal for the introduction of velocity perturbations due to a subhalo.

Each particle in a stream obeys the equation
\begin{equation}\label{linearevol}
\Delta\bs{\theta} = \btheta-\btheta_0 = (\boldsymbol{\Omega}-\boldsymbol{\Omega}_0)t_s +\Delta\bs{\theta}_\mathrm{init}= \Delta\bs{\Omega}_\mathrm{init}t_s+\Delta\bs{\theta}_\mathrm{init},
\end{equation}
where the subscript $0$ denotes the coordinates of the progenitor, $\Delta$ denotes separation between the progenitor and a stream particle and the subscript `$\mathrm{init}$'' denotes the separation between progenitor and particle at release and $t_s$ is the time since the particle was stripped. This means $\Delta\bs{\Omega}$ is the separation in frequencies between a stream particle and the progenitor, which is constant after the particle has been stripped, and $\Delta\bs{\theta}$ is the separation in angles between a stream particle and the progenitor which grows linearly in time at a rate $\Delta\bs{\Omega}$.

A long thin stream is characterised by a vector $\bs{n}$ along which the particles lie in both angle and frequency space. Under the assumption of an isotropic action distribution this vector is the principal eigenvector of the Hessian matrix $D_{ij}=\partial^2 H/\partial J_i\partial J_j$. Note that if the frequencies are functions solely of the Hamiltonian (as in the Kepler case) $\bs{n}$ is aligned with the stream particle frequency vectors and the stream is well approximated by an orbit. We will return to this point later in Section~\ref{Sect::analytic}.

\cite{Bovy2014} and \cite{Sanders2014} introduced a model for each tail of the stream (leading or trailing) in angle-frequency space that was an elongated Gaussian in the frequency offset from the progenitor and a isotropic Gaussian in initial angle offset from the progenitor. By estimating some stripping rate $p(t)$ the angle-frequency distribution could be calculated at all times. Through the introduction of a linear transformation from position-velocity space to angle-frequency space in the neighbourhood of the stream, \cite{Bovy2014} demonstrated that the stream track in configuration space can be computed quickly and the model could be rapidly sampled at any time.

The angle-frequency model of a stream can be simply extended to include the effects of a subhalo impact. Under the assumption that the subhalo imparts an instantaneous velocity kick to the stream particles, we can calculate the change to the angles and frequencies for small velocity kicks as
\begin{equation}\label{eq:kickfreqangle}
\begin{split}
\delta \boldsymbol{\Omega}^g &\approx \frac{\upartial \boldsymbol{\Omega}}{\upartial \bs{v}}\Big|_{\bs{x}}\cdot \delta\bs{v}^g,\\
\delta \bs{\theta}^g &\approx \frac{\upartial \bs{\theta}}{\upartial \bs{v}}\Big|_{\bs{x}}\cdot \delta\bs{v}^g.
\end{split}
\end{equation}
This is a good approximation as the transformation between $(\bs{x},\bs{v})$ and $(\boldsymbol{\Omega},\bs{\theta})$ is close to linear for the small velocity kicks from subhaloes.

The stream stretches along the direction $\bs{n}$ and in each tail the frequency distribution is very narrow such that, at a fixed time, the locations of the particles in the stream are well approximated by a single angle coordinate $\theta_{||}=\btheta\cdot\bs{n}$. We assume that the kicks are functions of this single variable. Similarly, the frequency distribution can be described by the single frequency coordinate $\Omega_{||}=\boldsymbol{\Omega}\cdot\bs{n}$. This assumption is equivalent to assuming the stream is very cold as we are ignoring any spatial extent of the stream perpendicular to the streaming direction. The distribution perpendicular to these narrow distributions are described using the angle and frequency coordinates
\begin{equation}
\begin{split}
\Delta\boldsymbol{\theta}_\perp &= \Delta\boldsymbol{\theta}-\Delta\theta_{||}\bs{n}\\
\Delta\boldsymbol{\Omega}_\perp &= \Delta\boldsymbol{\Omega}-\Delta\Omega_{||}\bs{n}.
\end{split}
\end{equation}
If the kick occurred a time $t_g$ ago the angles and frequencies of a stream particle are given by
\begin{equation}\label{linearkickevol}
\begin{split}
\Delta\bs{\theta} &= \Delta\bs{\theta}_{\rm init}+\Delta\boldsymbol{\Omega}_{\rm init}(t_s-t_g)+\delta\bs{\theta}^g+(\Delta\boldsymbol{\Omega}_{\rm init}+\delta\boldsymbol{\Omega}^g)t_g , \\&= \Delta\bs{\theta}_{\rm init}+\Delta\boldsymbol{\Omega}_{\rm init}t_s+\delta\bs{\theta}^g+\delta\boldsymbol{\Omega}^g t_g,\\
\Delta\boldsymbol{\Omega} &= \Delta\boldsymbol{\Omega}_{\rm init}+\delta\boldsymbol{\Omega}^g.
\end{split}
\end{equation}
Note that these kicks are only valid if $t_s>t_g$ i.e. the particle was in the stream at the kick time, otherwise the applied kicks to the angles and frequencies are zero.
The particle moves with its initial frequency separation until the kick at which point the particle continues to move at the initial frequency separation plus the frequency kicks.

The matrices $\frac{\upartial \bs{\theta}}{\upartial \bs{v}}\Big|_{\bs{x}}$ and $\frac{\upartial \bs{\Omega}}{\upartial \bs{v}}\Big|_{\bs{x}}$ must be calculated numerically. However, we will demonstrate that there are several approximate analytic relations that hold well for the inspected simulation.

\section{Simulation}\label{Sect::Simulation}

\begin{figure}
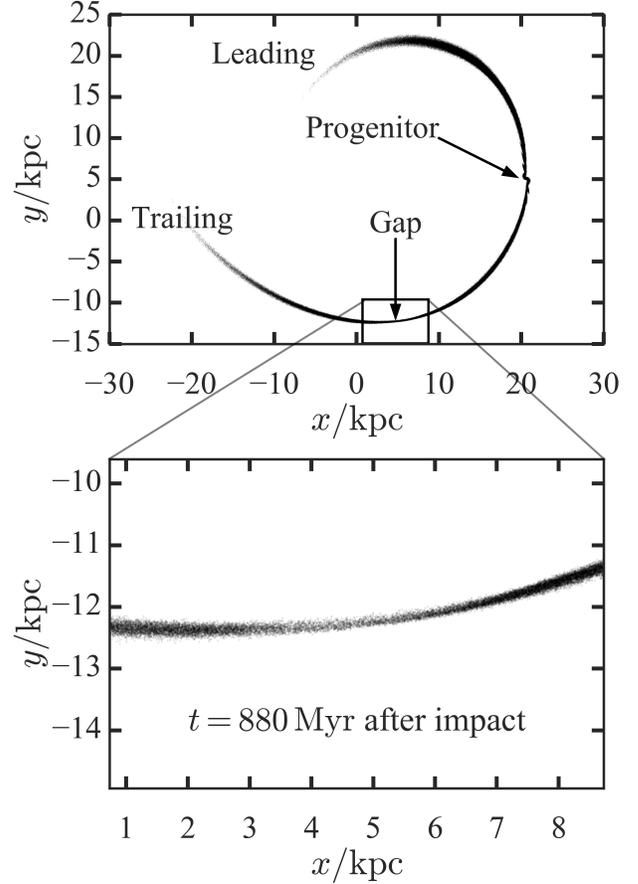

$$\includegraphics[width=\columnwidth]{{{plots/fig4_tilted_gap_morphology}}}$$
\caption{
Real-space distribution of stream and gap zoom-in for the $10^8M_\odot$ subhalo fly-by $880\Myr$ after impact.
}
\label{GapDiagram}
\end{figure}

\begin{table}
\caption{Parameters used for the $N$-body simulation described in Section~\ref{Sect::Simulation}. The parameters above the divide refer to the stream progenitor system whilst those below refer to the dark-matter subhalo. The progenitor of the stream follows a King profile, whilst the subhalo follows a Plummer profile.}
\begin{tabular}{lll}
\hline
Stream
&Mass&$10^5 M_\odot$\\
&Core radius&$13\pc$\\
&$W_0$&5\\
&Particle number&$10^6$\\
&Smoothing length&$1\pc$\\
&Initial position&$(30,0,0)\kpc$ \\
&Initial velocity&$(0, 105.75, 105.75)\kms$ \\
\hline
Subhalo&Mass&$10^7 M_\odot,10^8 M_\odot$\\
&Scale radius&$250 \pc,625 \pc$\\
&Impact parameter&$0$\\
&Velocity at impact&$200\kms\,\perp$ to stream velocity \\
&Impact time&$10\Gyr$ after stream progenitor released\\
&Insertion time&$100\Myr$ before impact\\
&Removal time&$100\Myr$ after impact\\
\hline
\end{tabular}
\label{SimParamsTable}
\end{table}

\begin{figure*}
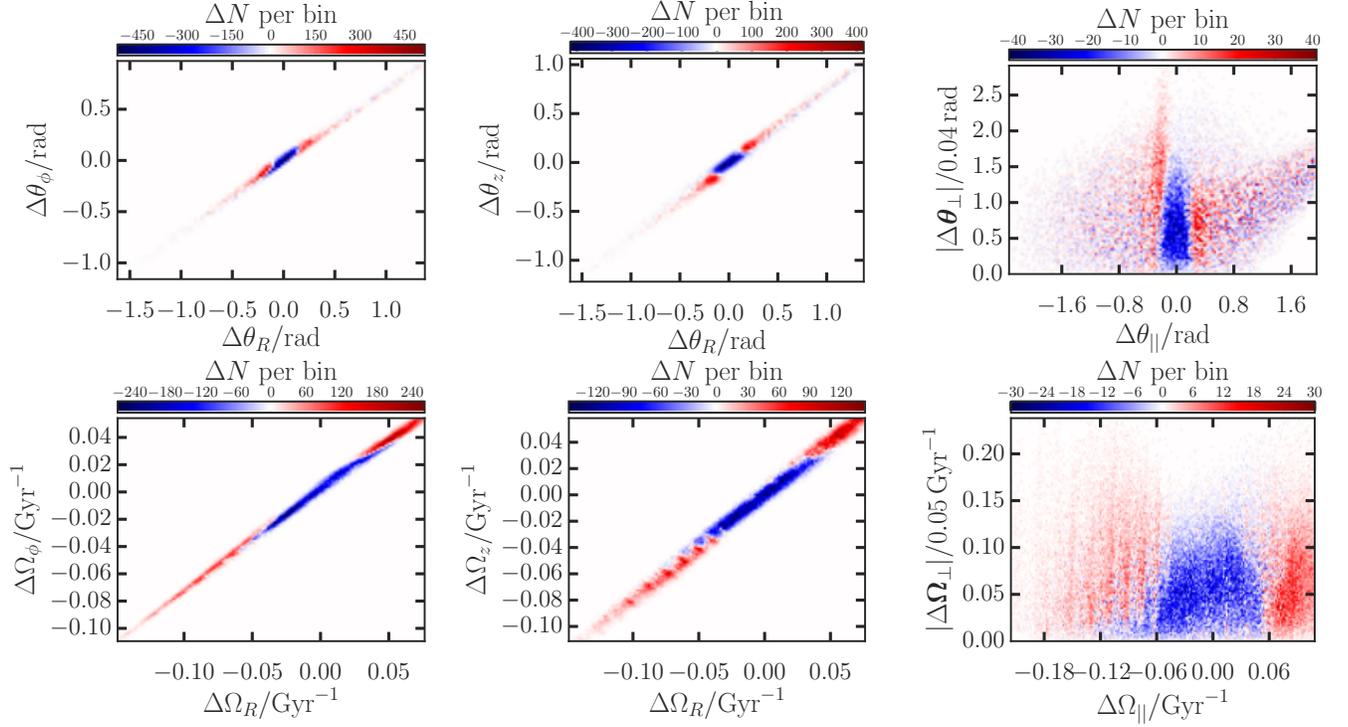

$$\includegraphics[width=\textwidth]{{{plots/fig5_tilted_diff_map}}}$$
\caption{
Difference histograms between the unperturbed and perturbed streams at $t=0.88\Gyr$ after impact: the \textbf{top row} shows the distribution in angles and \textbf{bottom row} the frequencies. The right panels show the distributions in the parallel-perpendicular space where `parallel' is the distance along the stream direction and `perpendicular' is the Euclidean distance from the stream direction vector. Zero-points correspond to the gap centre as described in Sec.~\protect\ref{Sec::ActAngFreq}. The bin size for the angles is $0.03\rad$ ($0.0012\rad$ for the perpendicular direction) and $0.015\Gyr^{-1}$ for the frequencies ($7.5\times10^{-4}\Gyr^{-1}$ for the perpendicular direction).
}
\label{tilted_diff_map}
\end{figure*}
To investigate the formation of stream gaps in actions, angles and frequencies we ran two $N$-body simulations. The simulations were run using \textsc{Gadget-3} which is an improved version of \textsc{Gadget-2} \citep{Gadget2}. The parameters of the simulation are summarised in Table~\ref{SimParamsTable}. We first generated a stream by disrupting a King profile with $W_0=5$, a mass of $10^5 M_\odot$ and a core radius of $r_c=13\pc$. The cluster was modelled with $10^6$ particles and a smoothing length of $1\pc$.
The galactic potential was chosen to be a logarithmic halo of the form in equation~\eqref{potential} with $V_c = 220\kms$ and $q = 0.9$. The cluster was placed on an eccentric orbit with a pericentre of $15\kpc$ and an apocentre of $30\kpc$. It was released around apocentre at a position of $(30,0,0)\kpc$ with an initial velocity of $(0, 105.75, 105.75)\kms$ such that the orbital period of the progenitor was $\sim650\Myr$. The simulation was run for $10\Gyr$ after which a long cold stream of total length $\sim 300^\circ$ was generated. This stream was then used in two simulations where it was impacted by two different Plummer subhaloes: a large subhalo with a mass of $10^8 M_\odot$ and a scale radius of $r_s = 625 \pc$, and a small subhalo with a mass of $10^7 M_\odot$ and $r_s = 250 \pc$. The impact point was in the trailing arm of the stream when it was near pericentre with an impact parameter of $b=0$ and a velocity which was perpendicular to the stream plane at the point of impact with a magnitude of $200\kms$. The subhalo was inserted into the simulation $100\Myr$ before impact and removed $100\Myr$ after impact to remove the possibility of multiple interactions with the stream. The stream-subhalo interaction time is of the order of $r_s/|\bs{w}-\bs{v}|$ where $\bs{v}$ is the velocity of the gap centre at impact. For the considered interaction, this expression gives $\sim 1.5\Myr$ which is significantly smaller than the time the subhalo is in the simulation. Each simulation was then evolved for a further $5\Gyr$.  In Figure~\ref{GapDiagram} we show the full stream in the $(x,y)$ plane and a zoom-in of the gap centre at $880\Myr$ after impact for the large subhalo. It is this simulation that we inspect throughout the paper and we have used the smaller subhalo simulation for validation.

This section is split into several subsections. In Section~\ref{Sec::ActAngFreq} we project the $N$-body simulation of the perturbed stream into action, angle and frequency coordinates. We then compute the angle and frequency kicks from Section~\ref{Sec::Formalism_angfreq} and compare to the perturbed and unperturbed simulations in Section~\ref{Sec::AngFreqKicks}. In Section~\ref{Sec::PerturbModel} we perturb the unperturbed snapshot using the numerically computed kicks and compare with the simulation. We develop an analytic expression for the frequency kicks in Section~\ref{Sect::analytic}. In Section~\ref{Sect::StreamGrowth} we discuss the stages of stream growth in angle space. In Section~\ref{Sect::anggapamp} we show how the angle kicks are related to the gap size in angle space and finally in Section~\ref{Sect::GapSize} show how the angle gap size is related to the spatial gap size.

\subsection{Angle \& frequency structure}\label{Sec::ActAngFreq}
Here we investigate the angle and frequency structure of the stream. We take the snapshot of the stream at $880\Myr$ after impact and compute the angles and frequencies for particles in the trailing stream tail (defined by a $\sim180^\circ$-long segment in azimuthal angle). We also perform the same calculation for the stream evolved without a subhalo flyby. In Fig.~\ref{tilted_diff_map} we show histograms of the difference in the density in angle and frequency between the perturbed and unperturbed simulation. We plot all quantities with respect to the coordinates of the gap centre which is defined by taking the unperturbed stream at the time of impact, computing the median velocity of the stream particles within $10\pc$ of the impact point, and then integrating the orbit starting at the impact point with this velocity. In the angle and frequency distributions there is a clear under-density along the stream direction, and we observe that the spreads in $|\boldsymbol{\theta}_\perp|$ and $|\boldsymbol{\Omega}_\perp|$ are significantly smaller than $\theta_{||}$ and $\Omega_{||}$ justifying our modelling assumption. In Appendix~\ref{Appendix::Analytic} we plot similar distributions in action space which exhibit very similar features.

\begin{figure}
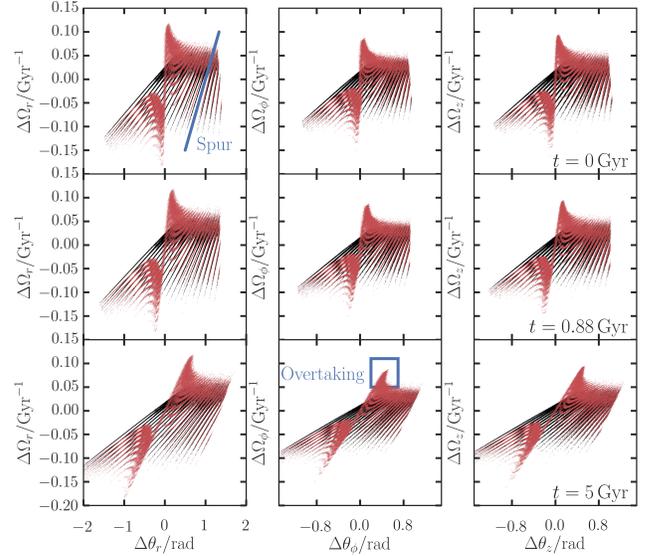

$$\includegraphics[width=\columnwidth]{{{plots/fig6_ang_freq_dists}}}$$
\caption{
Frequency-angle distributions for unperturbed (black) and perturbed (red) trailing tail at three different times after the subhalo impact. The three rows show the frequency and angle distributions with respect to the centre of the gap at three different times which are written in the right column. The three columns from left to right show the radial angle against radial frequency, the azimuthal angle against azimuthal frequency and vertical angle against vertical frequency.
% The bottom row shows the normalized 1D angle distributions at the three times with the dotted line corresponding to the $0\Gyr$ case, the dashed corresponding to $0.88\Gyr$ and the solid line corresponding to $5\Gyr$.
In the top right panel we show a blue line that lies along one of the `spurs' in the unperturbed model. In the bottom middle panel we highlight those kicked particles that are now overtaking other stream particles.
}
\label{angfreq_dist}
\end{figure}

In Figure~\ref{angfreq_dist} we plot the angles and frequencies for the unperturbed and perturbed trailing stream distributions. The unperturbed simulation consists of a spur (marked in blue in Figure~\ref{angfreq_dist}) due to each stripping event. The material stripped in each event forms an approximately vertical line in this space before gradually twisting clockwise as the differential frequency effects take hold. The material on the far left was stripped earliest. We see that the subhalo produces the expected S-shape which also twists clockwise in time due to the differential frequency effects. At the impact time ($t=0$) there is a clear gap in frequency but only a relatively modest gap in angle. However, the twisting naturally produces a larger gap in the angles. The twisting also causes some perturbed material to overtake the unperturbed material (highlighted by the blue box in Figure~\ref{angfreq_dist}).

\subsection{Angle and frequency kicks}\label{Sec::AngFreqKicks}
In Section~\ref{Sec::Formalism_angfreq} we described how the angle and frequency kicks due to a subhalo can be computed under a linear approximation from the velocity kicks. Here we use the expressions and compare the resulting kicks with those measured from the $N$-body simulation.

\begin{figure*}
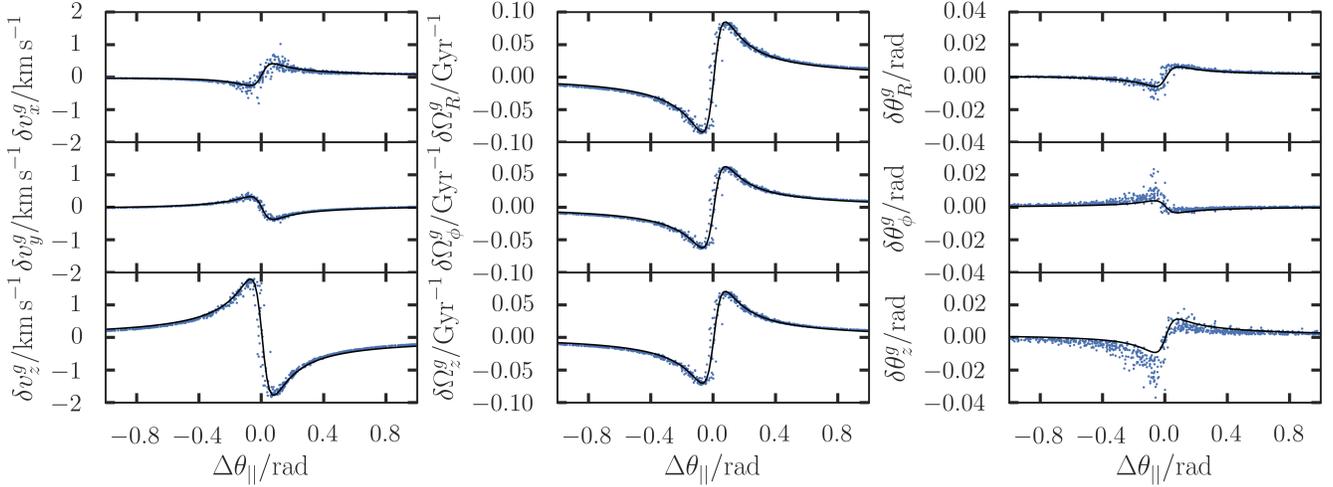

$$\includegraphics[width=\textwidth]{{{plots/fig7_tilted_angfreq_kicks}}}$$
\caption{
Angle and frequency kicks: the blue points are a random sample of $1000$ particles from the stream and show the kicks found from the simulations whilst the black lines show those calculated under the impulse approximation. The functional form for the kicks in both the angles and frequencies are very similar to that in velocities. Also, the angle and frequency kicks computed from the simulation match the numerical results well.
}
\label{kicks_nzero}
\end{figure*}

We begin by taking the snapshot of the unperturbed stream at the impact time and form a stream track in angle-frequency space by fitting a spline to the stream particles. We then compute the vector $\bs{n}$ from this track and at each point along the stream track compute $\theta_{||}=\btheta\cdot\bs{n}$ as well as $\delta\bs{\Omega}^g(\theta_{||})$ and $\delta\bs{\theta}^g(\theta_{||})$ from equation~\eqref{eq:kickfreqangle} by finite differencing in velocity at fixed position. We show the resultant kicks in Fig.~\ref{kicks_nzero}. We also show the velocity, angle and frequency kicks computed from the simulation. The perturbed velocities at impact are found by integrating the simulation backwards from $880\Myr$ after impact. In the case of the frequency kicks we simply difference the perturbed and unperturbed simulation. In the case of the angles we calculate the angles and frequencies of the perturbed simulation snapshot at $880\Myr$ after impact and rewind to the impact point before differencing with the unperturbed snapshot.

The fact the velocity kicks match well demonstrates that the assumptions made to derive the formulae in Section~\ref{Sect::Formalism} are valid whilst the use of the impulse approximation and the linear approximation from equation~\eqref{eq:kickfreqangle} is validated by the match of the frequency kicks. Interestingly the amplitude of the velocity kicks found from the numerical calculation overpredicts that in the simulation far from the impact centre. This corresponds to the difference observed in Fig.~\ref{curved_straight} between computing the kicks using the full stream track and subhalo orbit (`full', green dashed line) and assuming the relative velocity between each stream particle and subhalo is fixed during the fly-by (`curved', black).

The angle kicks are perhaps a less satisfying match than the velocity and frequency kicks. This may be due to the assumption of the impulse approximation or the neglected stream dispersion. We performed several checks of the accuracy of the angle computation: we measured the fluctuations in the angles about a straight line using a series of samples from a $300\Gyr$ segment of the progenitor's orbit. If computed exactly, the angles should lie along a straight line so deviations about this are due to numerical error. We also compared a snapshot evolved in angle-frequency space (i.e. using equation~\eqref{angleactdef}) with a later simulation snapshot projected into angle-frequency space. These tests show that the angles are computed to an accuracy of $\sim5\times10^{-4}\rad$ such that the discrepancy observed here cannot be due to errors in the computation.

\begin{figure*}
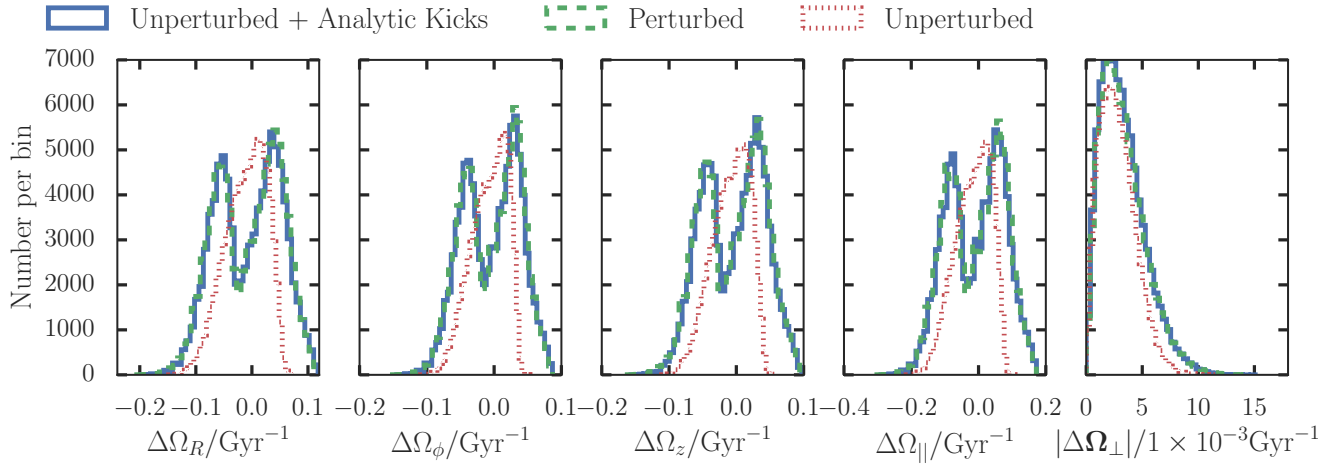

$$\includegraphics[width=\textwidth]{{{plots/fig8_tilted_model_perturbed}}}$$
\caption{
Frequency distributions for the perturbed and unperturbed stream model. The red histograms show the unperturbed stream, the green show the perturbed stream and the blue show the unperturbed stream kicked in frequencies using equations~\protect\eqref{eq:kickfreqangle}. The left three panels show the distributions in the orbital frequencies with respect to the gap centre. The fourth panel shows the frequency distribution along the stream direction $\hat{\boldsymbol{n}}$ whilst the fifth panel shows the distribution perpendicular to this.
}
\label{model_nzero}
\end{figure*}

\subsection{Model-simulation comparison}\label{Sec::PerturbModel}
With the angle and frequency kicks satisfactorily calculated we proceed to take the unperturbed simulation, apply the kicks and compare with the perturbed simulation at later times.

In Fig.~\ref{model_nzero} we show the frequency distributions of the unperturbed simulation, the perturbed simulation and our model constructed by perturbing the frequencies of the unperturbed simulation. The match is very good and there is a clear gap in all the components of the frequencies. We also show the parallel frequency distribution which exhibits a clear gap and the perpendicular frequency distribution which is slightly broader in the perturbed case than the unperturbed case reflecting the increase in velocity dispersion of the stream due to the subhalo.

\begin{figure}[h]
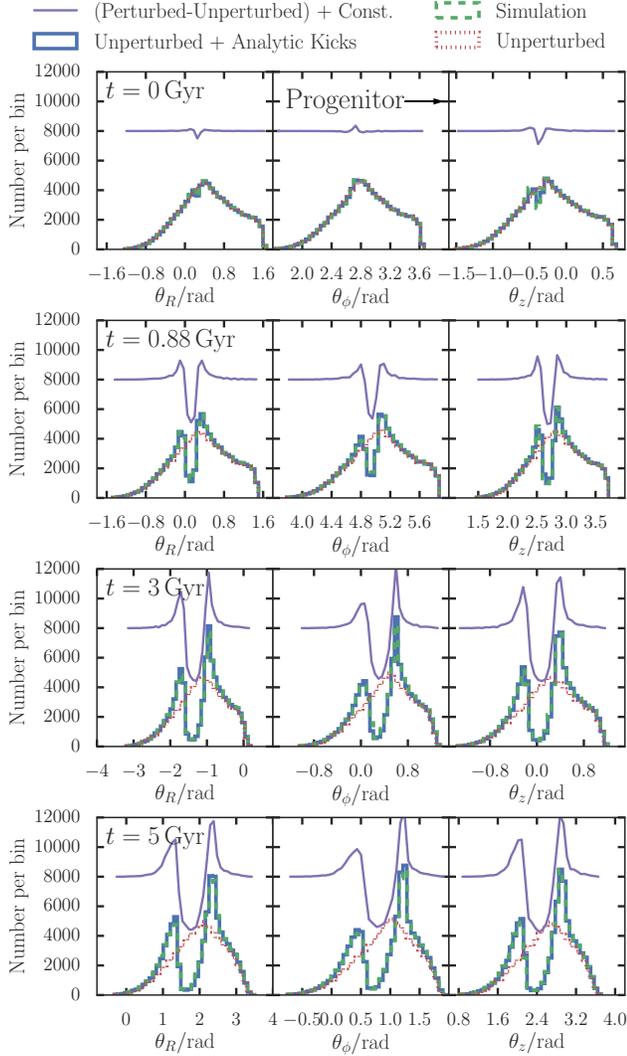

$$\includegraphics[width=\columnwidth]{{{plots/fig9_ang_later_times}}}$$
\caption{
Unperturbed, perturbed and model stream angle distributions: the red histograms show the unperturbed stream, the green show the perturbed stream and the blue show the unperturbed stream kicked in angle and frequency space and evolved. The purple lines show the difference between the perturbed and unperturbed streams offset by some arbitrary amount. The four rows show $0,0.88,3$ and $5\Gyr$ after impact.
}
\label{model_severaltimes}
\end{figure}

In Fig.~\ref{model_severaltimes} we plot the unperturbed and perturbed stream in angle space at four different times ($0\Gyr,0.88\Gyr,3\Gyr$ and $5\Gyr$ after impact) along with the model constructed by perturbing the unperturbed stream with the impulse approximation using equation~\eqref{eq:kickfreqangle}. Additionally we show the difference between the perturbed and unperturbed angle distributions. We see at all times there is a gap in $\theta_z$ and $\theta_R$ with an overdensity at $0\Gyr$ after impact in $\theta_\phi$. The gap grows in time in all angles, and our model matches the distributions well at all times, despite the fact that the angle kicks do not match the simulation particularly well (see Fig.~\ref{kicks_nzero}). This must be because the angle distributions are a combination of the frequency and angle kicks and the frequency kicks dominate at late times.

\subsection{Analytic approximations}\label{Sect::analytic}
We have seen that the angle distributions formed from kicking the unperturbed simulation match the perturbed simulation well at late times suggesting that the angle kicks are insignificant and dominated by the frequency kicks. To understand this, we now develop some understanding of the relative magnitude of the angle and frequency kicks by inspecting some cases where we can calculate these quantities analytically. In Appendix~\ref{Appendix::Analytic} we also present some approximate analytic results for the action kicks due to a subhalo fly-by.

\begin{figure}
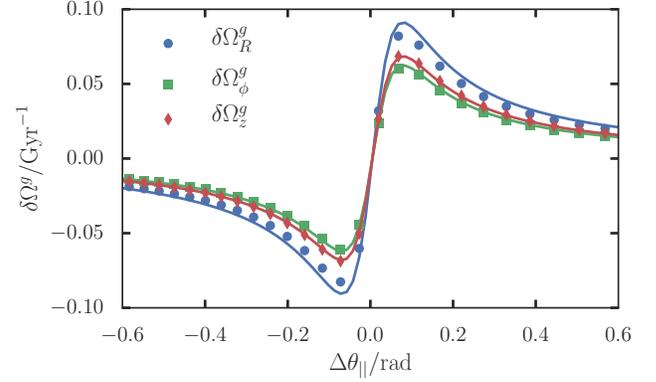

$$\includegraphics[width=\columnwidth]{{{plots/fig10_analytic_freq}}}$$
\caption{
Analytic frequency kick approximation. The lines show the analytic relationship for the flattened logarithmic potential from equation~\protect\eqref{dOm_an} and the symbols show the numerical calculation.
}
\label{analytic_freq}
\end{figure}

For scale-free potentials of the form $\Phi\propto r^\alpha$ \cite{WilliamsEvans2014} demonstrated that the Hamiltonian in action-space is well approximated by
\begin{equation}
H(\bs{J}) \propto (J_R+B L)^\beta,
\end{equation}
where $\beta=2\alpha/(2+\alpha)$, $B$ is a constant and $L$ is the angular momentum. They also suggest that in scale-free flattened axisymmetric potentials of the form $\Phi\propto (R^2+(z/q)^2)^{\alpha/2}$ the Hamiltonian is well approximated by
\begin{equation}
H(\bs{J}) \propto (J_R+B J_\phi+CJ_z)^\beta,
\end{equation}
where $B$ and $C$ are constants.
In these potentials the frequencies depend on the actions solely through the Hamiltonian as $\Omega_i\propto H^{(\beta-1)/\beta}$ so we find that
\begin{equation}
\frac{\delta\Omega_i^g}{\Omega_i}=\frac{\beta-1}{\beta}\frac{\delta H}{H}.
\end{equation}
Note in the case of the harmonic oscillator $\beta=1$ and the right-hand side vanishes as the frequencies are independent of energy. This equation can also be derived when considering the change in the energy and hence azimuthal frequency of a circular orbit. The analogous approximate expression for the Hamiltonian in the scale-free logarithmic potential is
\begin{equation}
H(\bs{J}) = V_c^2\log(J_R+B J_\phi+C J_z)),
\label{eqn::approxH}
\end{equation}
such that
\begin{equation}
\frac{\delta\Omega_i^g}{\Omega_i}=-\frac{\delta H}{V_c^2}.
\label{dOm_an}
\end{equation}
In the case of a subhalo flyby $\delta H^g=\bs{v}\cdot\delta\bs{v}^g$. We plot this approximation for the simulated stream in Fig.~\ref{analytic_freq}. The match is very good which suggests that the approximate Hamiltonian of equation~\eqref{eqn::approxH} is a very close approximation of the true Hamiltonian for the region of action space we are exploring here.

From our investigations there do not appear to be any particularly neat or understandable analytic expressions we can derive for the changes to the angle coordinates. What follows are some loose arguments to get a handle on the amplitude of the angle kicks relative to the frequency kicks. We inspect the harmonic oscillator (which is also appropriate for small oscillations about a circular orbit and perpendicular to the plane). In this case, the angle coordinate $\theta_i$ is given by
\begin{equation}
\tan\theta_i = \frac{\Omega_i x_i}{v_i},
\end{equation}
where $x_i$ is the position coordinate relative to the minimum, $v_i$ is the velocity coordinate. For the angle kicks we find that
\begin{equation}
\delta\theta^g_i\approx \frac{\tan\theta_i}{1+\tan^2\theta_i}\Big(\frac{\delta\Omega^g_i}{\Omega_i}-\frac{\delta v_i}{v_i}\Big),
\end{equation}
For a true harmonic oscillator the change in frequency is identically zero such that this expression can be reduced to
\begin{equation}
\delta\theta^g_i\approx -\frac{\delta v_i}{\sqrt{E}}\sin\theta_i = -\frac{1}{2}\frac{\delta v_i}{v_i}\sin 2\theta_i,
\end{equation}
where $E$ is the energy of the oscillation. We see that near a turning point ($v_i\approx0$) the angle kicks are largest and of order $\delta v_i/\sqrt{E}$ and near the midpoints ($x_i\approx0$) the angle kicks are approximately zero. As expected the magnitude of the angle kicks are phase-dependent. Using equation~\eqref{dOm_an} we see that
\begin{equation}
\frac{\delta\theta_i^g}{\delta\Omega_i^g}\approx\Omega_i^{-1}\Big(\frac{V_c}{|\bs{v}_i|}\Big)^2\sin 2\theta_i,
\end{equation}
where we have neglected geometric factors. The final two terms are of order unity such that these rather loose arguments convince us that this ratio is approximately the period of the oscillation in dimension $i$. From inspection of Fig.~\ref{kicks_nzero} we see that the ratio of the angle to frequency kicks is of order $0.1\Gyr$ which corresponds approximately to $1/\Omega$ as $\Omega_i\approx10\Gyr^{-1}$. From inspection of the simulation it appears $|\delta\theta^g_i|\approx\delta\Omega^g_i/\Omega_i$ within factors of $3$ for $|\Delta\theta_{||}|<1\rad$. Therefore, we have ascertained that the angle kicks are dominated by the frequency kicks after approximately one period.

The study of \cite{ErkalBelokurov2015} showed that gaps only begin to form in the stream after $\sim$ one radial period so an observed gap is always in the regime where the frequency kicks are dominating and when modelling we can essentially neglect the angle kicks.

\subsection{Stages of stream growth}\label{Sect::StreamGrowth}
\cite{ErkalBelokurov2015} discussed the three phases of stream gap formation in the limit that the stream itself is on a circular orbit. They found that on short time-scales (less than a radial period) an overdensity formed as the particles were scattered onto epicyclic orbits that brought them towards the gap centre. This was dubbed the compression phase. After approximately a radial period the stream enters an expansion phase where the gap begins growing until the more strongly perturbed material starts to overtake the more weakly perturbed material and caustics form. During the expansion phase the stream gap grows linearly in time whilst during the caustic phase the growth rate slows to $t^{1/2}$. In this section we investigate and discuss how this picture relates to the formalism presented here.

\begin{figure}
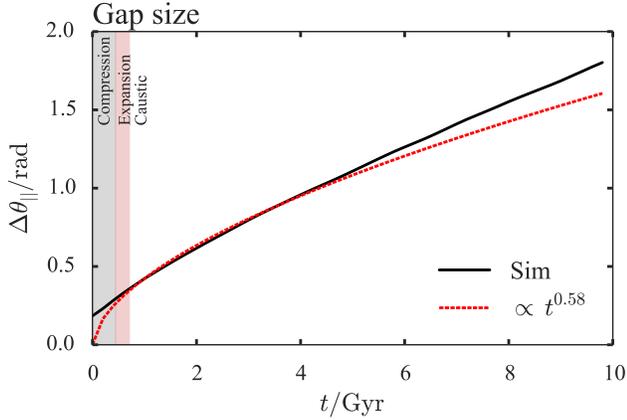

$$\includegraphics[width=\columnwidth]{{{plots/fig11_tilted_gap_size}}}$$
\caption{
Gap size in parallel angle as a function of time. The zero-point in time is the subhalo impact time. The black line shows the gap measured from the simulation and the dashed red line shows the best fitting power-law line for $t<5\Gyr$. We also mark on the three phases of the gap growth from \protect\cite{ErkalBelokurov2015}.
}
\label{gap_size}
\end{figure}

In Fig.~\ref{gap_size} we plot the size of the gap in parallel angle as a function of time for the simulated stream. This is defined as the difference between the parallel angles of the points where the perturbed and unperturbed densities are equal. We produce a smooth difference histogram using a Gaussian kernel density estimation with a bandwidth of $\Gamma = 0.04\rad\Gyr^{-1}t+0.01\rad$ for $t<1\Gyr$ and $0.05\rad$ otherwise and then fit a spline to the resulting distribution. With this choice of kernel, there are $\sim1000$ particles within one kernel standard deviation at the gap minimum and $\sim5000$ at the maxima. We also show a best-fit power-law line ($\Delta\theta_{||}\propto t^{0.58}$) for the segment $t<5\Gyr$ as well as the time-scales for the different stages of gap growth. The compression phase lasts for approximately a radial period. The time-scale on which the caustic becomes important is computed from equation (26) of \cite{ErkalBelokurov2015}. In this example the expansion phase only lasts for $\sim200\Myr$ but lowering the subhalo mass to $10^7M_\odot$ increases its length to $\sim600\Myr$. We note that there is a gap in parallel angle at the impact time ($t=0$). This does not correspond to a spatial gap but is due to the angle kicks from the subhalo that have shifted the stars to slightly different orbital phases.

The initial angle kick appears to dominate on a time-scale of $\sim400\Myr$ after which the best-fit power-law is a much better match. This also appears to correlate with the end of the compression phase. Therefore, it seems that the compression phase is associated with the initial angle kicks. We can understand this by considering a simple example of a stream on a radial orbit. Here the parallel angle is purely the radial angle. The shape of the angle kicks will be qualitatively similar to those in Fig.~\ref{kicks_nzero} with the particles in front of the gap given a positive kick and those behind given a negative kick. At a fixed position increasing the radial angle means we are now closer to apocentre so the apocentre has moved inwards and the particle must move more slowly. Likewise decreasing the radial angle moves apocentre further out and the particle must move faster. Therefore, the particles behind the gap move faster than those in front and an overdensity forms.

We use the formulae from \cite{ErkalBelokurov2015} to calculate the time-scale on which the caustic begins to form. In this case the expansion phase is very short and so the subsequent evolution should go with the square-root of time. We see, however, that this is only approximately true up to $\sim4\Gyr$ (the best-fitting power law is $t^{0.58}$) after which the growth rate is faster than this. We will return to this point later in Section~\ref{Sect::Variation} and only mention here that it is due to the gap forming on an already growing underlying stream.

\subsection{Amplitude of the angle kicks}\label{Sect::anggapamp}

\begin{figure}
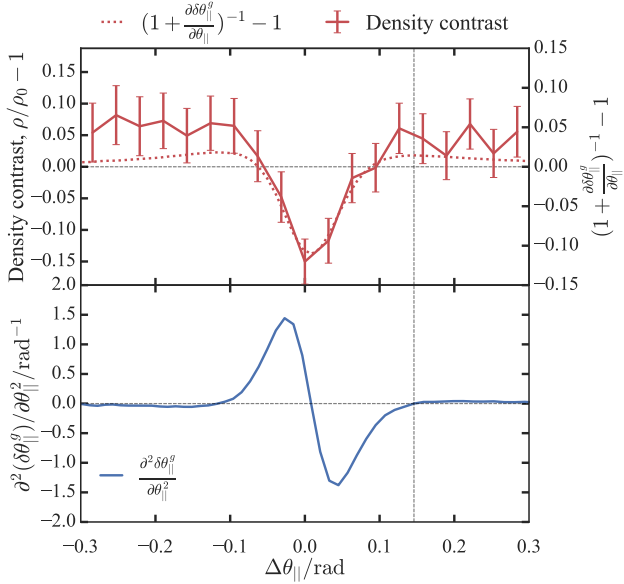

$$\includegraphics[width=\columnwidth]{{{plots/fig12_d2dthetadtheta2}}}$$
\caption{
Derivatives of the parallel angle kicks: the top panel shows the expression from equation~\eqref{deriv} in dashed red alongside the density contrast (the perturbed density in parallel angle space divided by the unperturbed density minus $1$) in solid red. The grey lines show the zero line and the approximate location of the peak density contrast. The bottom panel shows the second derivative of the parallel angle kicks with respect to the parallel angle.
}
\label{secondderiv}
\end{figure}

It is intriguing that the size of the initial gap in parallel angles ($\sim0.2\rad$) is significantly larger than the peak of the angle kicks ($\sim0.02\rad$) shown in Fig.~\ref{kicks_nzero}. However, one must consider the collective effects of a series of particles being kicked in order to compute the gap size. For instance, if the unperturbed parallel angle distribution $N_0(\theta_{||})$ were a uniform distribution the perturbed distribution $N(\theta_{||})$ would be given by
\begin{equation}
N(\theta_{||})=\int {\rm d}\theta'_{||}\,N_0(\theta'_{||})\delta(\theta_{||}-\theta'_{||}-\delta\theta^g_{||}(\theta'_{||}))\propto(1+\frac{\partial\delta\theta^g_{||}}{\partial\theta_{||}})^{-1}.
\label{UniformAssump}
\end{equation}
The gap size is then related to the zeros of $\partial\delta\theta^g_{||}/\partial\theta_{||}$
and the peaks in the perturbed distribution are at the zeros of the second derivative
$
\partial^2\delta\theta^g_{||}/\partial\theta^{2}_{||}.
$
In Fig.~\ref{secondderiv} we show the second derivative computed for our example as well as the quantity
\begin{equation}
(1+\frac{\partial\delta\theta^g_{||}}{\partial\theta_{||}})^{-1}-1.
\label{deriv}
\end{equation}

The second derivative is zero around $\delta\theta_{||}\approx0.14\rad$ and equation~\eqref{deriv} crosses zero around $\delta\theta_{||}\approx0.1\rad$. These correspond well to the observed initial amplitude of the parallel-angle gap size $(\sim0.2\rad)$. In Fig.~\ref{secondderiv} we also show the density contrast computed from the simulations which correlates well with equation~\eqref{deriv}.

We can now understand why the initial gap size in parallel angle is much greater than the magnitude of the kicks in parallel angle. Fig.~\ref{kicks_nzero} shows how the angle kicks depend upon $\Delta\theta_{||}$ and we see that the kicks peak at small $|\Delta\theta_{||}|$ with a long tail at larger $|\Delta\theta_{||}|$. Additionally, the range of $\Delta\theta_{||}$ over which the kicks are important is larger than the magnitude of angle kicks. When considering the gap that forms we have to consider the collective effects of many particles over a large range in $|\Delta\theta_{||}|$ being kicked a small amount and the gap that forms is larger than perhaps expected. If the angle kicks were large compared to the range in $\Delta\theta_{||}$ over which the kicks are important the gap size would be approximately the amplitude of the angle kicks. Therefore, we conclude that there is not a one-to-one correspondence between the amplitude of the angle kicks and the initial parallel angle gap size, and the latter can be significantly greater than the former.

\subsection{Physical and parallel angle gap size}\label{Sect::GapSize}

\begin{figure}
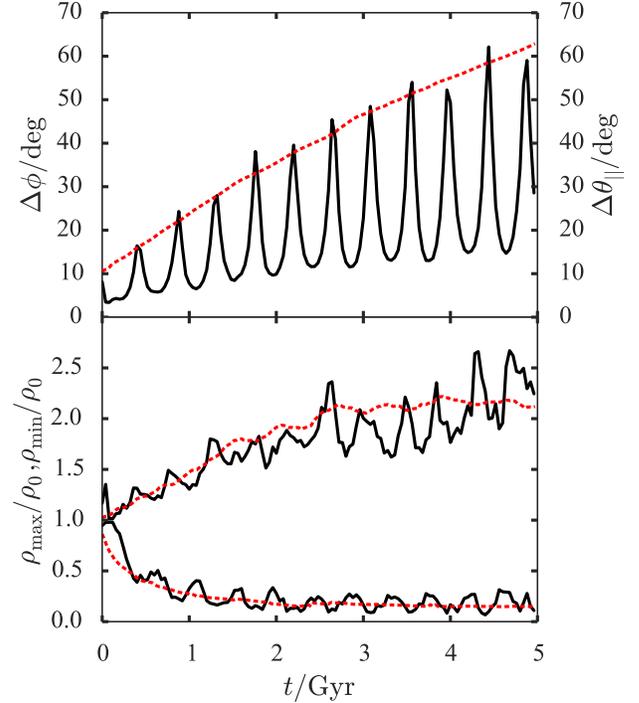

$$\includegraphics[width=\columnwidth]{{{plots/fig13_real_vs_angle_properties}}}$$
\caption{
Gap properties in real and angle space as a function of time. The zero-point of time is the subhalo impact time. In the \textbf{top panel} we show the gap size in the azimuthal angle of a plane fitted through the stream in solid black and the gap size in parallel angle in dashed red. In the \textbf{bottom panel} we show the minimum and maximum density contrast in these two spaces.
}
\label{real_v_angle}
\end{figure}
Our discussion so far has focussed on the size of the gap in parallel angle. Whilst this is an appropriate space for understanding general stream morphology it is awkward to see the relationship with the observed stream properties. We briefly investigate this by projecting the model back into configuration space. For this purpose we use the publicly-available torus code from \cite{BinneyMcMillan2015} (\href{https://github.com/PaulMcMillan-Astro/Torus}{https://github.com/PaulMcMillan-Astro/Torus}). We require a different numerical algorithm to map from angle-action space back into configuration space as the algorithm from \cite{SandersBinney2014} transforms from configuration space to angle-action space.

We construct eight tori on the corners of an action-space box that encompasses the full perturbed distributions and interpolate the frequencies to find the actions of the perturbed particles. As described in \cite{BinneyMcMillan2015} we interpolate the Fourier coefficients of the tori in action-space to construct a torus of the required actions for each particle and request the $(\bs{x},\bs{v})$ corresponding to the angles of the particle. Each torus is constructed with a relative tolerance in the actions of $2\times10^{-6}$ and recovers the positions and velocities of the particles to an absolute accuracy of $\sim0.01\kpc$ and $\sim0.1\kms$ respectively. This accuracy is sufficient for inspecting the density and width of the gap which is of order a few $\kpc$ after a $\Gyr$.

We construct a plane that minimises the perpendicular spread of the stream and compute the size of the gap in the azimuthal angle in the plane in the same way as in parallel angle space. We plot both the azimuthal gap size and the parallel angle gap size as a function of time in Fig.~\ref{real_v_angle}. As expected the azimuthal gap size oscillates but it is well enveloped by the parallel angle gap size and it appears that the average azimuthal gap size correlates well with the parallel angle gap size. We note that the amplitude of the oscillations of the azimuthal gap size can be estimated by considering conservation of angular momentum. If two particles in the stream had identical angular momentum their azimuthal separation $\Delta\phi\sim r^{-2}$ resulting in a gap size which varies by a factor of $(r_\mathrm{max}/r_\mathrm{min})^2$ over the orbit. For our simulation means the azimuthal separation fluctuates by a factor $\sim 4$, which agrees well with Fig.~\ref{real_v_angle}.

We also plot the minimum and maximum density contrasts (defined as perturbed divided by unperturbed density). The azimuthal density contrast oscillates about the parallel angle density contrast, and both plateau at late times ($t>4\Gyr$). We can see the gap forms immediately in parallel angle whilst in azimuthal angle there is a small peak due to the compression phase. \cite{ErkalBelokurov2015} showed that the minimum density contrast falls as $1/t$ at large times, and the maximum grows as $t$ at small times and like $1/(1-at)$ at large times with caustics forming at $t=1/a$. These properties are well reproduced by our simulation. Continuing the approximate modelling of equation~\eqref{UniformAssump} and including the assumption that the particles all move at the same frequency we find that the density contrast at later times is proportional to
\begin{equation}
(1+\frac{\partial\delta\Omega^g_{||}}{\partial\theta_{||}}t)^{-1}-1.
\label{densitycontrast}
\end{equation}
This expression is much like those derived by \cite{ErkalBelokurov2015} in the circular-orbit limit. For a uniform unperturbed angle distribution the minimum density contrast occurs at $\Delta\theta_{||}=0$ and so falls as $1/(1+Ct)$ where $C$ is a constant. The maximum density occurs when $|1+(\partial\delta\Omega^g_{||}/\partial\theta_{||})t|$ is minimised. At early times this expression is non-zero for all $\Delta\theta_{||}$ whilst at later times the expression is zero at some $\Delta\theta_{||}$ such that equation~\eqref{densitycontrast} diverges and the caustic behaviour of \cite{ErkalBelokurov2015} is recovered. We note that in reality the minimum density contrast in Fig.~\ref{real_v_angle} falls more slowly than $1/t$ at large times and actually reaches a constant.

The discrepancy between our simulation and the picture of \cite{ErkalBelokurov2015} seems to be due to our stream having a non-zero velocity dispersion. This means each stripping event produces stream particles with a range of energies and upstream (i.e. closer to the progenitor) particles with large energy differences from the progenitor can pass through the stream and fill in a downstream (i.e. further from the progenitor) gap. Inspection of Fig.~\ref{angfreq_dist} shows that if the parallel frequency kicks are greater than the spread in frequencies of the stream a clear gap will form and will not be significantly filled in by upstream particles. This is essentially a condition on the velocity kicks being greater than the velocity dispersion of the stream. In this picture, the unperturbed stream density also plays an important role as if there are many particles upstream from the gap then the density contrast will be rapidly washed out. The stripping rate in the simulation we are examining decreases with time (see Section~\ref{Sect::FULL}) so the filling-in effect is less severe in our example than it potentially could be. The models of \cite{ErkalBelokurov2015} and \cite{ErkalBelokurov2015b} allow for a non-uniform unperturbed density distribution but with no stream velocity dispersion the stars in their model cannot pass through the stream to affect a gap.

\section{Varying the impact properties}\label{Sect::Variation}
We have demonstrated that we are able to adequately model the impact of a subhalo on a stream and the subsequent evolution. With this machinery in place we are able to rapidly simulate the effects of any subhalo fly-by. In this section we will discuss the differences in the subsequent stream structure as a function of the subhalo mass and the subhalo impact location. We repeat the exercise performed in Section~\ref{Sect::Simulation} of simulating the subhalo fly-by on a simulated unperturbed stream and inspecting the subsequent evolution.

\subsection{Varying the subhalo mass}
\begin{figure*}
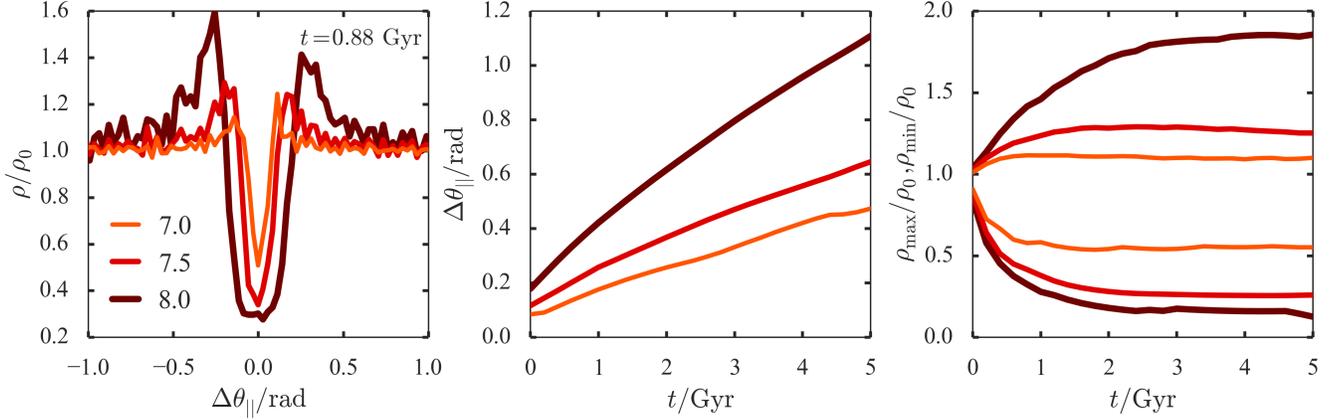

$$\includegraphics[width=\textwidth]{{{plots/fig14_different_masses}}}$$
\caption{
Properties of the stream as a function of subhalo mass. The thick brown line corresponds to $10^8M_\odot$, the medium red line $10^{7.5}M_\odot$ and the thin orange line $10^7M_\odot$. The left panel shows the difference between the perturbed and unperturbed simulations in parallel angle $880\Myr$ after impact, the middle panel shows the gap size as a function of time and the right panel shows the minimum and maximum density contrast as a function of time.
}
\label{varying_mass}
\end{figure*}

First, we investigate the properties of the stream as a function of subhalo mass. We adopt the scaling relation between the mass and scale radius of the subhalo Plummer sphere of
\begin{equation}
M\propto r_s^{2.5}.
\end{equation}
This relation produces a series of subhalos that fall within the distribution of the maximum circular velocity against total (tidal) mass for the haloes from the Via Lactea II simulations \citep{Diemand2008}. There is an uncertainty in the power-law slope of $\sim0.7$. Keeping all other properties of the subhalo fly-by the same we simulate the effects of a $10^7M_\odot$ and $10^{7.5}M_\odot$ subhalo. In Figure~\ref{varying_mass} we show the gap profile $880\Myr$ after impact, the gap size as a function of time and the minimum and maximum density contrast as a function of time. Again we have used the smoothing kernel mentioned in Section~\ref{Sect::StreamGrowth} with an additional factor in the bandwidth depending on the subhalo mass $M$ of $0.5\log_{10}(M/10^6M_\odot)$. As expected from the results of \cite{ErkalBelokurov2015}, the gap size is deepest and grows fastest for the highest mass subhalo. In all cases the minimum density contrast plateaus in time and the value to which the density contrast plateaus decreases with increasing mass. The ratio between the asymptotic values of the density contrast for the $10^8M_\odot$ and the $10^7M_\odot$ case is observed to be $\sim4$.

\cite{ErkalBelokurov2015} showed that, for stream particles on a circular orbit, the central density contrast falls to zero like $1/t$ with the constant of proportionality scaling as $M/r_s^2$. For all subhalo masses we observe that the density contrast plateaus in time and, as discussed in the previous section, is due to a combination of the stream density and the stream energy distribution. The time at which the density contrast reaches its plateau value decreases with decreasing mass.

\subsection{Varying the impact geometry}
\begin{figure}
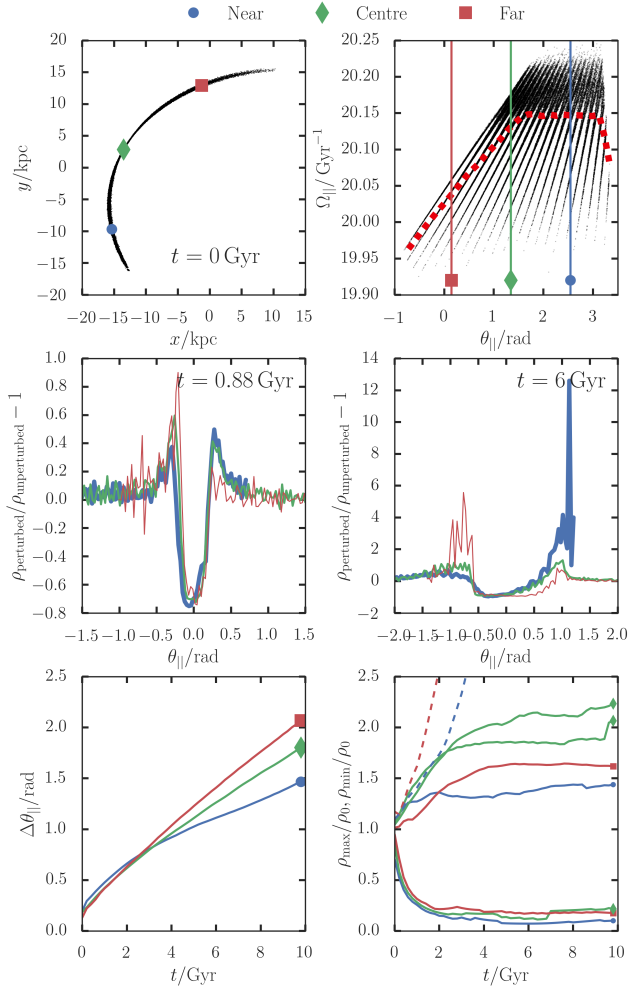

$$\includegraphics[width=\columnwidth]{{{plots/fig15_different_strike_points}}}$$
\caption{
Properties of the stream as a function of the location of the impact point. \textbf{Top row}: The top left panel shows the stream in the $(x,y)$ plane along with three symbols showing the different impact points at the impact time (the impact nearest the progenitor is marked with a blue circle, the impact furthest from the progenitor is marked with a red square and an impact between these two extremes is marked by a green diamond). The top right panel shows the location of these strikes in parallel angle (the vertical lines are coloured and marked with the symbols from the top left panel) as well as the distribution of the stream in parallel frequencies and angles. \textbf{Middle row}: The middle two panels show the relative density contrast $\mathrm{d}N/\mathrm{d}\theta_{||} = \rho_\mathrm{perturbed}/\rho_\mathrm{unperturbed}-1$ in parallel angle space for two separate times ($880\Myr$ and $6\Gyr$ after impact). The `near' impact corresponds to the broadest line and the `far' impact to the narrowest. \textbf{Bottom row}: The bottom two panels show the gap size in parallel angle and the minimum and maximum density contrast as a function of time with the dashed lines showing the divergent peak at low unperturbed density. The lines are coloured and marked to correspond with the top left panel.
}
\label{varying_geometry}
\end{figure}

Now we investigate how the stream changes as a function of where along the stream the impact occurs. In addition to the inspected geometry of Section~\ref{Sect::Simulation} we simulate two other impacts at varying distance from the progenitor as shown in the top left panel of Figure~\ref{varying_geometry}. The impact close to the progenitor occurs at ($[x,y,z]=[-15.3,-9.8,9.3]\,\mathrm{kpc},[v_x,v_y,v_z]=[41.9,-183.0,152.2]\,\mathrm{km\,s}^{-1}$) and the one further from the progenitor at ($[x,y,z]=[-1.2,12.9,-10.8]\,\mathrm{kpc},\,[v_x,v_y,v_z]=[-239.8,-100.6,87.0]\,\mathrm{km\,s}^{-1}$). Again, we fix all other parameters of the impact to those in the original simulation, including the absolute velocity of the subhalo (note that fixing the relative velocity between the subhalo and impact point in the stream does not change the conclusions drawn here). The near impact is in the regime where the mean parallel frequency is approximately constant with the parallel angle whilst the far impact is in the regime where the mean frequency is increasing linearly with the parallel angle. In this far regime the stream is well ordered in energies, whilst in the near regime the stream particles have not had sufficient time to order themselves by energy. We plot the difference between the perturbed and unperturbed simulations in parallel angle for two times. The density contrast at large negative parallel angle for the \emph{far} case is very large at late times as the number density in the unperturbed simulation at these large separations is low. A similar effect is seen at large positive parallel angle for the \emph{near} case but here it should be noted that we have not included the effect of more stars entering the stream as they are stripped from the progenitor. In the lower panels we show the gap size and the minimum/maximum density contrast as a function of time. The gap size in all three cases is very similar for $t<3\Gyr$ but they diverge at larger times as the far impact gap size increases faster than the near impact. We can understand this as in the far case the gap is forming on an already growing stream as the underlying stream structure is well ordered. However, in the near case the stream is mixed so the underlying stream is growing more slowly. In all cases the minimum density plateaus to a similar value. Note that we have shown two lines for the maximum density contrast for the two peaks on either side of the gap. In both the far and near cases one of the peaks diverges rapidly and correlates with the observations in the middle panels.

\section{A full generative perturbed stream model}\label{Sect::FULL}
Up until now we have used our $N$-body simulation as our unperturbed stream model to which we have applied the kicks. We now produce a full model of the perturbed stream that starts from a model for the unperturbed stream. For the latter, we use the model presented in both \cite{Bovy2014} and \cite{Sanders2014}. This is a fully generative model that models the expected stream structure in angle--frequency space and projects this structure into the space of observables. In angle--frequency space a model is expressed in full generality as
\begin{equation}
\begin{split}
p(&\boldsymbol{\Omega},\btheta,t_s) =\\
 & p\big(\Delta\btheta_\mathrm{init}=\Delta\btheta-\Delta\boldsymbol{\Omega}t_s|\boldsymbol{\Omega}_\mathrm{init},t_s)\,p(\Delta\boldsymbol{\Omega}_\mathrm{init}=\Delta\boldsymbol{\Omega}|t_s)\,p(t_s)\,.
\end{split}
\end{equation}
(see equation \eqref{linearevol} which defines $\Delta\btheta$). The stripping time $t_s$ is the time since the particle was stripped from the progenitor. The specific model investigated by \cite{Bovy2014} and \cite{Sanders2014} simplified this to
\begin{equation}\label{eq:simplemodel}
p(\boldsymbol{\Omega},\btheta,t_s) = p(\Delta\boldsymbol{\Omega}_\mathrm{init})\, p(\Delta\btheta_\mathrm{init})\,p(t_s)\,,
\end{equation}
by assuming that the distribution of frequency offsets $\boldsymbol{\Omega}_\mathrm{init}$ and angle offsets $\btheta_\mathrm{init}$ is independent of stripping time $t_s$. In Section~\ref{Sect::modelmods} we discuss how this is \emph{not} a good assumption for the simulated stream considered in this paper and how we modify the model to account for this. However, as discussed by \cite{Bovy2014}, the simple model above is still useful, as it allows the location of the stream in frequency--angle and configuration space to be efficiently calculated, both at the present time and at the time of impact.

In the model of \cite{Bovy2014} and \cite{Sanders2014}, $p(\Delta\boldsymbol{\Omega}_\mathrm{init})$ is a bimodal distribution, because there are peaks at positive and negative $\Delta\boldsymbol{\Omega}_\mathrm{init}$ corresponding to the leading and trailing tails. Here we focus on modelling the trailing tail so we only consider negative $\Delta\boldsymbol{\Omega}_\mathrm{init}$. The distribution $p(\Delta\boldsymbol{\Omega}_\mathrm{init})$ is chosen to be as in \cite{Bovy2014}. This distribution is approximately Gaussian with axes that approximately align with the eigenvalues of the Hessian matrix $D_{ij}=\partial^2H/\partial J_i\partial J_j$ \citep{BinneyTremaine}. Note that they do not exactly align, because the anisotropic action distribution of the tidal debris is taken into account. The angle distribution $p(\Delta\btheta_\mathrm{init})$ is a simple isotropic Gaussian. In the simplest model, the stripping rate $p(t_s)$ is taken as a uniform distribution up to some maximum time $t_d$---the disruption time---in \cite{Bovy2014}. For the uniform stripping rate, the location of the mean stream track in frequency--angle space can be computed analytically. To compute the transformation from angles and frequencies to Galactocentric Cartesian coordinates we interpolate the linear transformation computed for a series of points along the stream track as described in \cite{Bovy2014}.

The model for the perturbed stream can be obtained from the model of the unperturbed stream by computing the kicks at the time of impact and applying these as in equation~\eqref{linearkickevol}. To compute the kicks we approximate the stream as being one dimensional along $\theta_{||}$ with a point of closest approach $\theta_{||,c}$, which is a model parameter. We use the same method as discussed in Section~4.2 of \cite{Bovy2014} to compute the mean stream track and the derivatives $\frac{\upartial \boldsymbol{\Omega}}{\upartial \bs{v}}$ and $\frac{\upartial \bs{\theta}}{\upartial \bs{v}}$ along the track. Then we compute the velocity perturbation $\delta\bs{v}^g(\theta_{||})$ due to the subhalo along the track and propagate it to the kicks in $\delta \boldsymbol{\Omega}^g$ and $\delta \bs{\theta}^g$ as in equation~\eqref{eq:kickfreqangle}. These frequency and angle kicks then fully specify the subsequent evolution of the stream (see equation~\eqref{linearkickevol}).

From this model we can generate mock streams by sampling a stripping time $t_s$, an initial frequency separation $\Delta\boldsymbol{\Omega}_\mathrm{init}$ and an initial angle separation $\Delta\btheta_\mathrm{init}$ from the unperturbed model. We then perturb them with $\delta \boldsymbol{\Omega}^g$ and $\delta \bs{\theta}^g$ based on their $\theta_{||}$ at the time of impact. With these chosen, the current angles and frequencies are known (see equation~\eqref{linearkickevol}) and hence the observables can be computed.

We set the velocity-dispersion parameter ($\sigma_v$) of the model by scaling that obtained in \cite{Bovy2014} for a similar, but lower mass, stream. As shown in \cite{SandersBinney2013a} for instance the properties of the stream in angles and frequencies scales with the mass of the progenitor, $M$, as $M^{1/3}$.

\begin{table}
\caption{Generative model parameters: we list each parameter in the modified perturbed generative model from Section~\ref{Sect::FULL}. The top section lists the parameters for the simple unperturbed stream model. The middle section lists the parameters added for the modified unperturbed model of Section~\ref{Sect::modelmods}. The bottom section lists the parameters used to model the perturbation. Note the parameters of the potential have not been listed.}
\begin{tabular}{ll}
$(\bs{x},\bs{v})$&Progenitor position \& velocity now\\
$\sigma_v$&Progenitor velocity-dispersion parameter\\
$t_d$&Disruption time\\
\hline
$w$&Evaporation-to-pericentric stripping weight\\
$\mu_\mathrm{peri}$&Centre of material stripped at pericentre\\
$\sigma_\mathrm{peri}$&Dispersion of material stripped at pericentre\\
$\mu_\mathrm{evap}$&Centre of evaporated material\\
$\sigma_\mathrm{evap}$&Dispersion of evaporated material\\
\hline
$M$& Subhalo mass\\
$r_s$& Subhalo scale radius\\
$\bs{w}$& Subhalo velocity at impact\\
$b$& Impact parameter\\
$t_g$& Impact time\\
\end{tabular}
\label{Parameters}
\end{table}

\subsection{Modifications to model}\label{Sect::modelmods}
\begin{figure}
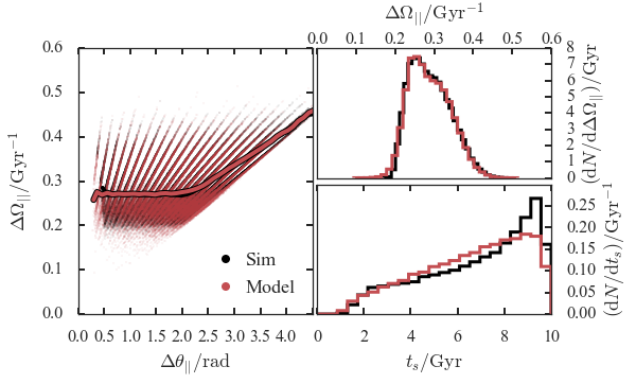

$$\includegraphics[width=\columnwidth]{{{plots/fig16_new_unperturbed_model}}}$$\\
\caption{Unperturbed stream model properties: the left panel shows the parallel angle separations against parallel frequency separations for particles from the simulation (in black) and samples of the generative model (in red). The solid line shows the mean parallel frequency separation for bins in parallel angle separation. The top right panel shows the parallel frequency separation distributions and the bottom right panel shows the stripping time distributions (where the colour-coding is the same as for the left panel). Larger stripping times mean the particles were stripped longer ago.}
\label{UnpertNEW}
\end{figure}

We found that the simple model described above did not completely describe the simulation (in particular the density along the stream near the gap) such that several modifications were necessary. Fig.~\ref{UnpertNEW} shows the results of our modifications, and illustrates the following discussion. Firstly, the stream distribution in $\Omega_{||}$ was not well described by a Gaussian, but instead seemed better fitted by two Gaussians. Inspection of the stream distribution in $\Delta\Omega_{||}$ and $\Delta\theta_{||}$ showed the stream consisted of material stripped in two separate ways: $\sim72\percent$ of the material is stripped around pericentric passage and this material forms a broad peak in $\Delta\Omega_{||}$ that is separated from the progenitor by $\mu_\mathrm{peri}=0.31\Gyr^{-1}$ with a width of $\sigma_\mathrm{peri}=0.05\Gyr^{-1}$. The other $\sim28\percent$ is material that is more continuously stripped, or that evaporates, from the progenitor. This material forms a narrower peak in $\Delta\Omega_{||}$ that is less separated from the progenitor at $\mu_\mathrm{evap}=0.25\Gyr^{-1}$ with a width of $\sigma_\mathrm{evap}=0.023\Gyr^{-1}$.

Additionally, inspecting the distribution of $\Delta\theta_{||}/\Delta\Omega_{||}\approx t_s$ showed that the stripping rate (averaged over a time longer than an orbital period of the progenitor) is not uniform, but instead decreases slowly over time (with decreasing stripping time), such that a particle is more likely to have been stripped long ago. We modelled this by making $p(t_s)\propto t_s$ for both the periodic stripping and for the more steady evaporation. We found that this modification was necessary to even approximately reproduce the density of the stream near the gap. Compared to the simple model above, the distribution $p(\Delta\Omega_{||},t_s)$ cannot be separated as $p(\Delta\Omega_{||})\,p(t_s)$ in this model and the distribution of parallel frequency separation and stripping time looks like
\begin{equation}
\begin{split}
p(\Delta\Omega_{||},t_s) \propto &(1-w)t_s\mathcal{N}(\Delta\Omega_{||}|\mu_\mathrm{peri},\sigma_\mathrm{peri})\\\times&\sum_n^{\mathrm{int}([t_d-t_R]/T_R)}\delta(t_s-nT_R-t_R)\\+&wt_s\mathcal{N}(\Delta\Omega_{||}|\mu_\mathrm{evap},\sigma_\mathrm{evap}),
\end{split}
\end{equation}
where $\mathcal{N}(X|\mu,\sigma)$ is a Gaussian distribution with mean $\mu$ and width $\sigma$. The parameter $w=0.28$ is the weight, $T_R$ is the radial period, $t_R$ is the time since last pericentric passage and $\delta(t_s-nT_R-t_R)$ is a delta function expressing stripping at each pericentre passage. Fig.~\ref{UnpertNEW} shows the improvements made to the model such that it matches the simulation well. The full set of parameters used in the model are listed in Table~\ref{Parameters}.

We use this modified model to generate mock stream particles. To compute the impact kicks in this model as a function of $\theta_{||}$, we use the stream track at the time of impact computed in the simple model with the uniform distribution of stripping times discussed in the previous section. This is because the mean track cannot be computed easily for the non-separable $p(\Delta\Omega_{||},t_s)$ distribution, but can be estimated using the uniform $p(t_s)$. This simplification should give a good estimate of the mean track as $p(t_s)$ mainly affects the density along the stream, but not its average location.

\subsection{Comparison to the $N$-body simulation}

\begin{figure}
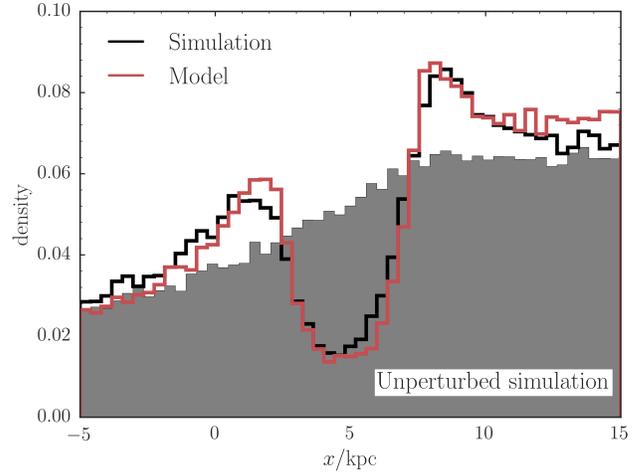

$$\includegraphics[width=\columnwidth]{{{plots/fig17_config_density}}}$$
\caption{Normalised number counts in a mock stream generated from the generative model described in the text (red) compared to that in the simulation (black) near the gap. The density of the unperturbed simulation is displayed in grey. The density in the simulation is well matched in the generative model.}
\label{config_density}
\end{figure}

We generate 100,000 mock stream particles from the modified generative model
for the $M=10^8\,M_\odot$ impact. The density along the stream in the
simulation and of the mock stream is displayed in
Figure~\ref{config_density}. The grey, filled histogram shows the
stream density in the unperturbed simulation for comparison. It is
clear that the simple generative model provides an excellent match to
the density in and near the gap in the simulation.

\begin{figure*}
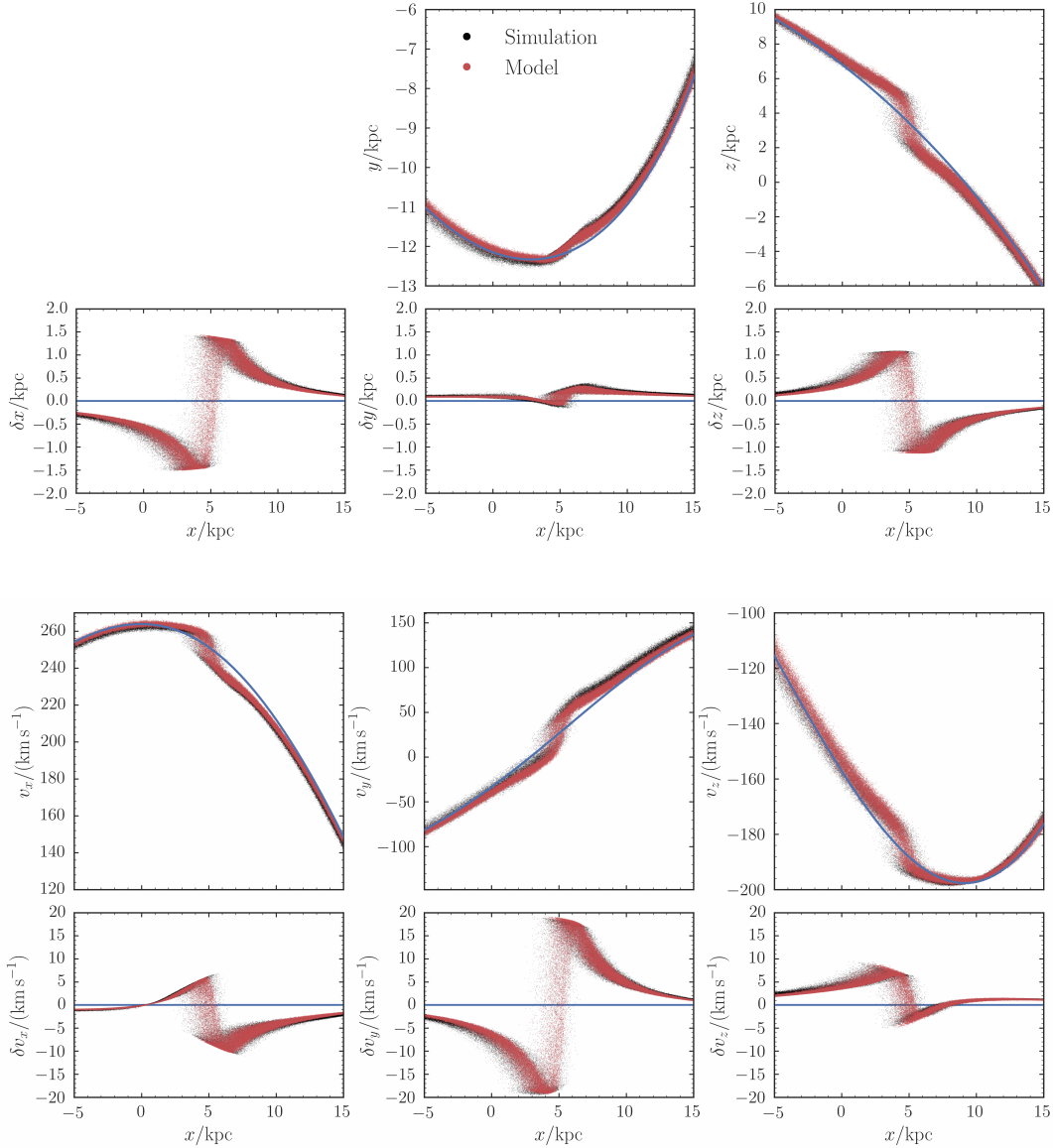

$$\includegraphics[width=0.8\textwidth]{{{plots/fig18a_config_position}}}$$\\
$$\includegraphics[width=0.8\textwidth]{{{plots/fig18b_config_velocity}}}$$
\caption{Comparison of the simulated stream (black) and the mock stream from the generative model (red) in configuration space. The blue line shows the unperturbed stream track. The bottom of each panel compares the difference in the phase--space position of each particle due to the subhalo perturbation in the simulation and the mock stream. The phase--space positions and the differences are plotted as a function of unperturbed $x$ position to exaggerate the perturbations; as a function of perturbed $x$ position, the stream closely follows the unperturbed track. The overall structure of the stream and in particular the phase--space offsets due to the subhalo perturbation are successfully modelled in the mock stream.}
\label{config_phasespace}
\end{figure*}

The simulated stream in configuration space near the gap is compared
to the mock stream in Figure~\ref{config_phasespace}. The mock stream
very closely follows the simulated stream. All quantities are plotted as a function of the unperturbed $x$ position. This exaggerates the size of the perturbations. The perturbation as a function of the perturbed $x$ position is smaller and in this space the perturbed stream more closely follows the unperturbed stream track. The bottom of each panel
shows the difference between the phase--space coordinates in the
perturbed and unperturbed simulation (black) and in the perturbed and
unperturbed mock stream (red). These are computed by comparing the
present coordinates of the same simulated/mock particles evolved with
and without the perturbation. That is, the perturbed and unperturbed
simulation start from the same exact initial conditions and we do the
same for the mock stream. It is clear from these bottom panels in
Figure~\ref{config_phasespace} that the perturbation due to the
subhalo impact is very well described by the generative model
described in this section.

\section{Conclusions}\label{Sect::Conclusions}
We have presented a framework for modelling the formation of gaps due to the perturbation of a dark matter halo on a cold thin stream formed from progenitors on eccentric orbits. The formalism uses the simple description of dynamics provided by the action, angle and frequency coordinates. Stream particles obey simple equations of motion in this space which can be altered to incorporate the effects of a dark matter subhalo fly-by. An unperturbed stream is stretched along a single direction, described by the parallel angle coordinate, in which a gap forms when the stream is impacted by a subhalo.

We presented an $N$-body simulation of a cold stream formed from a progenitor on an eccentric orbit in a scale-free flattened axisymmetric logarithmic potential. The stream was impacted by a $10^8M_\odot$ dark-matter subhalo and a gap formed in the stream. We analysed the formation of the gap in angle-frequency space and showed how the stream at all future times could be modelled using our framework. We found that generically the angle perturbation due to a fly-by is only important on time-scales less than a radial period. At later times the frequency perturbations are much more important and are simply related to the velocity perturbations in scale-free potentials such that the structure of the gap at late times is controlled by $\bs{v}\cdot\delta\bs{v}^g$.

We computed how the gap forms in the parallel angle (the angle along the stream) and showed that the distribution in this space can be simply related to the real-space properties of the gap. The evolution of the gap in angle-frequency space can be understood in the framework presented by \cite{ErkalBelokurov2015} and we verify that almost all the intuition developed in the circular orbit approach translates to results observed in the more realistic eccentric orbit case. At times greater than approximately $10$ orbital periods of the progenitor orbit ($\sim6\Gyr$) after the impact, we find several exceptions to the \cite{ErkalBelokurov2015} picture: the density of the gap plateaus in time as stream material fills in the growing gap and the stream gap growth rate depends on the position along the stream at which the subhalo impact occurs, with gaps furthest from the progenitor growing fastest. For times less than $10$ orbital periods of the progenitor orbit our results corroborate the \cite{ErkalBelokurov2015} picture.

We showed how the structure of the gap varies as a function of the mass of the subhalo and the location along the stream the impact occurs. We found that, as expected from the work of \cite{ErkalBelokurov2015}, the parallel angle gap grows more slowly in time for lower mass subhaloes. The density contrast of the gap minimum plateaus to a constant value after a few $\Gyr$ for all inspected subhalo masses but the depth of this plateau increases with mass. The more interesting result is that the growth rate of the gap depends on the location along the stream where the impact occurs. We found that near the progenitor where the particles are mixed in energy the gap grows more slowly than far from the progenitor where the particles are well ordered in energy and so the gap forms on an already growing stream.

We closed by presenting a fully generative model of a perturbed stream in Galactocentric Cartesian coordinates based on the model presented in \cite{Bovy2014} and \cite{Sanders2014}, and showed that the model matches the $N$-body simulation qualitatively very well. We found it necessary to modify the simple model by making the stripping rate increase in time, and including two populations of stripped particles: one population is stripped around pericentre with large energy shifts from the progenitor whilst the other population evaporates from the cluster continuously with smaller energy shifts.

Throughout the paper we have used a single simulation of a $10^8M_\odot$ subhalo impact. We have also analysed the same simulation with a $10^7M_\odot$ subhalo impact and as expected the amplitude of the effects observed in this paper are decreased but qualitatively the picture is very similar.

\subsection{Comments \& Outlook}

The formalism presented promises to be crucial for constraining the properties of a subhalo fly-by on a cold stream. It is anticipated that one would find potential gap candidates in star counts as has been demonstrated by \citet{CarlbergPal52012} and \citet{CarlbergGD12013} and then one would follow up with more detailed observations of the kinematic structure of the gap. \cite{ErkalBelokurov2015b} discussed what range of subhalo properties one could potentially constrain with a current and future observation quality. They found that the shape of the gap encodes the subhalo properties and showed that with ongoing surveys like the Dark Energy Survey and \textit{Gaia}, it will be possible to characterise dark subhaloes down to $10^7M_\odot$. This analysis was limited to streams on near-circular orbits but it is anticipated that many of the insights apply to more general stream geometries. We may yield stronger constraints using stream progenitors on eccentric orbits but this remains to be seen from future analysis. It is clear that the framework presented here allows all of the parameters of a subhalo impact to be varied efficiently for any stream, because all of the computationally-expensive steps can be pre-computed once the smooth stream model is known.

Whilst an underdensity in a stream is indicative of a subhalo fly-by, subhalo interactions are not the only cause of structure in a stream. \cite{Kuepper2010} have shown that overdensities naturally form in an eccentric stream due to epicyclic oscillations and this effect is seen most clearly near the progenitor. When expressing the model in angle-frequency coordinates we naturally include all the dynamics of the stream particles and the epicyclic overdensities are formed from the mapping from angle-frequency coordinates back to configuration space. We have shown in this paper that the effect of subhaloes significantly alters the structure in angle-frequency space and so will produce a qualitatively different signature to the epicyclic overdensities. Additionally, the rate at which the stream stars are stripped from the progenitor affects the density structure in the stream. Pericentric passage stripping events (which also occur when the stream progenitor is on an eccentric orbit) naturally produce substructure in the stream nearest the progenitor where the particles have yet to order in energy. In the models presented in \cite{Bovy2014} and \cite{Sanders2014} the stripping rate was assumed constant and here we have extended this approach to account for a variable stripping rate. In our simulation the subhalo struck the stream far from the progenitor where, even with a variable stripping rate, the unperturbed density was very smooth. However, for impacts close to the progenitor it may become more difficult to disentangle the effects of variable stripping with a subhalo impact. Fortunately for a long stream that has gone through many stripping events the stripping rate may be quantifiable due to its anticipated periodicity, but for shorter streams there may be considerable degeneracy. Finally, we comment that substructure in streams can be due to incorrect modelling of the background or variable extinction but these are not issues with the stream modelling so do not concern us here.

In this paper we have analysed the effect of a single subhalo fly-by. \cite{YoonJohnstonHogg} have predicted that a stream such as Palomar-5 would only experience a few $10^7$-$10^8M_\odot$ subhalo impacts but that there are many more much lower mass impacts. We have shown that a clear gap forms for the high mass impacts but it would be interesting to investigate the properties of the stream that has experienced many small mass subhalo impacts. In this regime we begin probing the mass spectrum of dark matter subhaloes. \citet{CarlbergGD12013},\citet{NganCarlberg2014} and \citet{ngan_etal_2015} have discussed the effects of many small impacts in simulations but in order to constrain the mass spectrum we require full models of perturbed streams.

The framework presented here is appropriate for when an interaction between the stream and a subhalo is short compared to the orbital time of the stream such that the impulse approximation is appropriate. Instead of a full perturbative analysis we were able to simply consider instantaneous changes to the stream properties. However, it is potentially a small starting step to developing a scheme to handle more general perturbations to streams. The assumption of a smooth analytic galaxy model is probably inappropriate, particularly over the timescales on which streams evolve. \cite{Bonaca2014} presented a study of analysing streams from the Via Lactea simulation under assumptions on the smoothness and evolution of the halo. They found that very massive subhalo encounters gave rise to minor overestimates in the halo mass whilst the time evolution of the potential produced biases of order $20\percent$ in the halo mass. Additionally, recent results suggest that the mass of the Large Magellanic Cloud is much larger than previously believed \citep{Kallivayalil2013, Penarrubia2016} which suggests that its potential must be included as a perturbation in the modelling of the Galaxy. However, a counter to this is that many of the observed streams are very simple cold structures suggesting that the approximation of a smooth halo for the Milky Way is valid. Nonetheless it is inevitable that, as data quality increase, one has to include perturbations on top of a smooth model to fit the known streams in the Galaxy and the angle-action framework is ideal for such an endeavour. We have demonstrated how a very simple perturbation can be included on top of a smooth angle-action model. These variables were originally introduced to handle perturbative solutions to the equations of the motion in the Solar system so seem highly appropriate for handling perturbations on Galactic scales.

The results presented here made use of the \texttt{galpy} package \citep{galpy}: the various methods for computing the velocity kicks outlined in Section~\ref{Sect::Formalism} and the methods for producing a full perturbed stream model as detailed in Section~\ref{Sect::FULL} are all implemented in the package. Additionally, the code used to generate all the figures in this paper is available in a \texttt{github} repository at \href{https://github.com/jobovy/streamgap-aa}{https://github.com/jobovy/streamgap-aa}.

In coming years the quality and quantity of data will provide us with detailed observations of many tidal streams. We anticipate that all of these streams have had a turbulent past full of impacts from dark subhaloes. Our grand goal is to reveal the history of each stream and extract the mass spectrum of dark matter subhaloes in the Milky Way. The formalism we have presented is the first step towards creating a very powerful tool for the job.

\section*{Acknowledgments}
JLS acknowledges the support of the Science and Technology Facilities
Council. JB thanks the Natural Sciences and Engineering Research
Council of Canada for financial support of this project. DE acknowledges the financial support from the ERC. The research leading to these
results has received funding from the European Research Council under
the European Union's Seventh Framework Programme (FP/2007-2013)/ERC
Grant Agreement no. 308024. We thank Raymond Carlberg and Paul McMillan for useful comments on a draft. Additionally, we thank the anonymous referee for a very thorough report that improved the presentation of some of the results.

\bibliography{bibliography}
\bibliographystyle{mn2e}

\appendix

\section{Analytic approximation for action kicks}\label{Appendix::Analytic}
In Section~\ref{Sect::analytic} we found a very accurate analytic approximation for the frequency kicks for scale-free potentials. In this appendix we attempt to find analogous expressions for the actions and angles.

\begin{figure*}
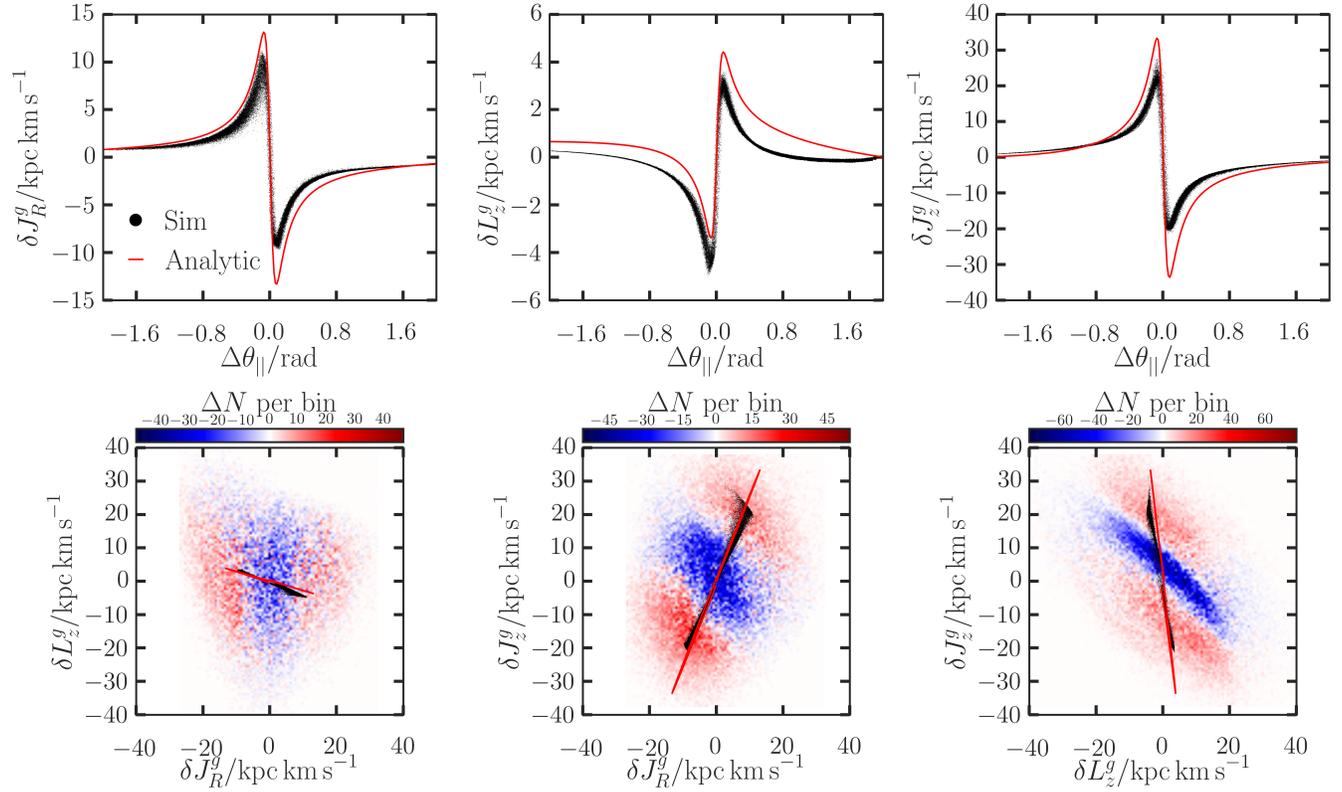

$$\includegraphics[width=\textwidth]{{{plots/figA1_tilted_analytic_action}}}$$
\caption{
Analytic action kick approximation. In the \textbf{top panels} the red shows the approximate analytic expressions from equation~\protect\eqref{dJ_an} and the black is computed from the simulation. In the \textbf{bottom panels} we overlay the action kick distributions on the difference histograms between the perturbed and unperturbed streams. The bin spacing for the difference histograms is $0.7\kpc\kms$. As with the angle and frequency difference distributions there is a clear gap formed at the impact centre. The analytic approximation explains the correlations observed in the simulation.
}
\label{aa_nz}
\end{figure*}

We approximate the changes in the actions\footnote{ Note that our unperturbed stream model is a function of the angles and frequencies such that the action kicks are unimportant for modelling purposes.} as
\begin{equation}
\begin{split}
\delta J^g_R &\approx \frac{\delta H^g - \Omega_\phi\delta L^g}{\Omega_R},\\
\delta J^g_\phi &\approx (\bs{r}\times\delta\bs{v}^g)_z,\\
\delta J^g_z &\approx \frac{v_z \delta v^g_z}{\Omega_z},
\end{split}
\label{dJ_an}
\end{equation}
where $\delta L = \bs{L}\cdot\delta\bs{L}^g/|\bs{L}|$ the change in the angular momentum. The change in $J_R$ comes from the spherical approximation and the change in $J_z$ from the assumption that it is a harmonic oscillator in the $z$ direction. We show these analytic approximations in Fig.~\ref{aa_nz} along with the kick distributions overlaid on difference histograms between the perturbed and unperturbed stream. As with the angle and frequency distributions in Figure~\ref{tilted_diff_map} we observe an under-density at the centre of the action distributions. The exact geometry of the under and over-densities is complex and clearly depends on the nature of the stream track and subhalo properties. Inspection of equation~\ref{dJ_an} shows that the direction of the action kicks depends on the arbitrary direction of the velocity kicks and the properties of the stream at impact. The analytic approximations recover the magnitude of the kicks within a factor of $\sim1.5$ but are not nearly as impressive as the frequency kick analytic expressions. These approximations work better for streams confined to the meridional plane. However, we see from the lower panels of Fig.~\ref{aa_nz} that the correlations between the action kicks are well recovered and explain the observed structure in the action difference plots.

\label{lastpage}
\end{document}